\documentclass{webofc}
\usepackage[varg]{txfonts}   
\begin{document}
\title{Probing the fission properties of neutron-rich actinides with the astrophysical $r$ process}
%
%

\author{\firstname{Nicole} \lastname{Vassh}\inst{1,6}\fnsep\thanks{\email{nvassh@nd.edu}} \and
        \firstname{Matthew} \lastname{Mumpower}\inst{2,6} \and
        \firstname{Trevor} \lastname{Sprouse}\inst{1,3} \and \firstname{Rebecca} \lastname{Surman}\inst{1,6} \and \\ \firstname{Ramona} \lastname{Vogt}\inst{4,5}
}

\institute{University of Notre Dame, Notre Dame, Indiana 46556, USA 
\and
           Theoretical Division, Los Alamos National Lab, Los Alamos, NM, 87545, USA
\and           
           Los Alamos Center for Space and Earth Science Student Fellow, Los Alamos National Lab, Los Alamos, NM, 87545, USA
\and
           Nuclear and Chemical Science Division, Lawrence Livermore National Laboratory, Livermore, CA 94551, USA
\and
           Department of Physics, University of California, Davis, CA 95616, USA
\and
           Joint Institute for Nuclear Astrophysics - Center for the Evolution of the Elements, USA
          }

\abstract{%
We review recent work examining the influence of fission in rapid neutron capture ($r$-process) nucleosynthesis which can take place in astrophysical environments. We briefly discuss the impact of uncertain fission barriers and fission rates on the population of heavy actinide species. We demonstrate the influence of the fission fragment distributions for neutron-rich nuclei and discuss currently available treatments, including recent macroscopic-microscopic calculations. We conclude by comparing our nucleosynthesis results directly with stellar data for metal-poor stars rich in $r$-process elements to consider whether fission plays a role in the so-called `universality' of $r$-process abundances observed from star to star.
}
\maketitle
\section{Introduction}\label{sec-1}
The potential to probe the origin(s) of the heaviest elements observed in nature has never been greater. Previously supernovae were thought to be responsible for the synthesis of these elements but most modern simulations suggest that core-collapse cases do not have the proper conditions to reach the heaviest observed nuclei (e.g. \cite{Bliss}). A currently accepted open window for supernovae to produce heavy species is a special case with high magnetic fields known as magneto-rotationally driven supernovae, however many simulations suggest this environment struggles to reach the actinide elements (e.g. \cite{Winteler,Mosta}). A long discussed possible site is that of neutron star merger events (e.g. \cite{Lattimer}) since neutron-rich material can be dynamically lifted from the surface of the neutron star(s) forming tidal tails during their approach or expelled more violently at the collision interface. Additionally an accretion disk can later form around the merger remnant and eject reprocessed material. In such conditions the rapid neutron capture process, $r$ process, has the possibility to synthesize the heaviest species known to occur in nature. 

The era of multi-messenger astronomy now permits unprecedented insights into astrophysical events such as mergers, as evidenced by LIGO/VIRGO's ability to direct the telescope community to perform detailed electromagnetic follow-up for merger events such as GW170817 (e.g. \cite{Cowperthwaite,Abbott,Kasen}). Although this event broke ground in providing direct evidence for the production of lanthanide elements, it remains unclear whether species heavier than this such as gold, platinum, and the actinides were produced. An interesting way to search for evidence that the heaviest species were produced is to look to identify signatures of fission in astrophysical observables since if the actinides have been reached then lighter species such as gold must have been synthesisized as well. However investigating fission in the $r$ process currently faces challenges from a general lack of understanding for the properties of the neutron-rich actinides. In this proceedings, we summarize the state of the nuclear data used for fission in $r$-process calculations and discuss currently identified potential signatures of fission within astrophysical environments.

\section{A brief discussion of fission rates for neutron-rich nuclei}\label{sec-2}

There are a limited number of available datasets for the fission reactions of neutron-rich nuclei (see \cite{VasshJPG,Eichler,Samuel,Goriely,KTUYapp} and references therein). Since the $r$ process operates in a regime far from current experimental data, simulation results must be reported as a function of the theoretical model applied. There are only a handful of fission barrier predictions for neutron-rich nuclei in the literature. All available models predict fission barriers to become low for many neutron-rich nuclei of importance to the $r$ process (see Figure 14 of \cite{VasshJPG} for a comparison of four publicly available barrier height sets). We note that although different barrier treatments imply differing reaches for ultimately how heavy of species the $r$ process can produce \cite{VasshJPG}, applications of rates determined from currently available fission barriers suggest that the $r$ process does not populate the predicted super heavy element island of stability around $Z=114$ and $N=184$ \cite{MattBDF}. For determining the abundances of $r$-process nuclei, currently calculations suggest neutron-induced and $\beta$-delayed fission to be of the greatest importance, with neutron-induced fission most commonly being predicted to be the means by which the reach of heavy element synthesis is terminated. However \cite{MattBDF} highlighted the importance of $\beta$-delayed fission during the late-times which set $r$-process abundances and for the first time predicted that multi-chance $\beta$-delayed fission can occur for neutron-rich nuclei. 

The limited availability of calculations for the fission properties of neutron-rich nuclei means prescriptions are sometimes not applied with self-consistent barriers for all fission reaction and decay channels. Since this can lead to nucleosynthesis results which may show features which are in actuality an artifact of such an inconsistency it is important to push toward a self-consistent implementation of fission properties in the $r$ process. To apply rates which have considered the same set of fission barriers in both the neutron-induced and $\beta$-delayed fission processes which dominate how the synthesis of heavy elements proceeds, we use the LANL suite of Hauser-Feshbach codes (see \cite{VasshJPG} and references therein). Using four distinct theoretical fission barrier sets, in \cite{VasshJPG} we examined exactly which nuclei most influence $r$-process predictions for the final abundances from an astrophysical process such as merger tidal-tail ejecta which can robustly produce fissioning species. We found many of these nuclei to be far from current experimental reach and therefore highlighted the need for progress in theoretical descriptions. However, since the fissioning nuclei of most importance are those populated as $\beta$-decay moves the general population of $r$-process nuclei toward stability \cite{VasshJPG,TrevorProc}, rather than those in the most neutron-rich regions near the dripline, dedicated future experiments may be able to close in on some of these species.

\begin{figure}[h!]
\centering
\includegraphics[scale=0.415]{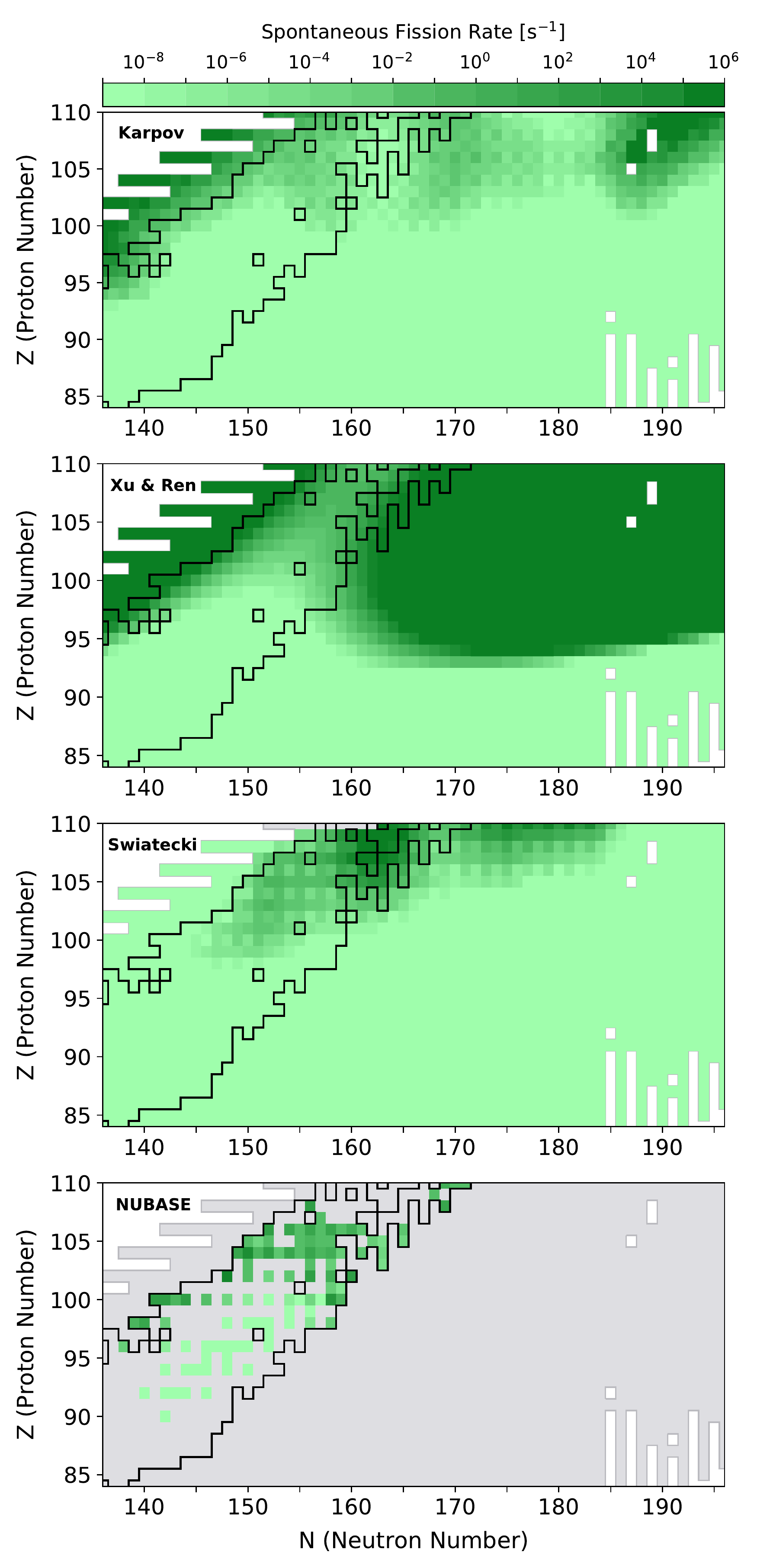}
\caption{Spontaneous fission rates assuming FRLDM barriers \cite{FRLDM} along with the phenomenological formula of Karpov {\it et al.} \cite{Karpov} (top panel) as compared to rates given by Xu $\&$ Ren \cite{XuRen} (second panel) and Swiatecki \cite{Swiatecki1955} (third panel) parameterizations. The bottom panel shows experimental rates given NUBASE2016 \cite{NUBASE2016} half-lives and branching ratios for comparison. The black outline shows the extent of NUBASE half-life data and the grey region shows the FRDM2012 \cite{FRDM2012} neutron dripline.}
\label{fig:spf}
\end{figure}

With respect to spontaneous fission, the rates available are often simple parameterized formulas fit to a limited set of data. This means such descriptions can disagree by orders of magnitude when their prescription is extrapolated into neutron-rich regions (see Fig.~\ref{fig:spf}). In our $r$-process calculations we apply spontaneous fission rates determined by the phenomenological equation of \cite{Karpov} which depends on fissility ($Z^2/A$) and barrier height. These spontaneous fission rates are very low until reaching $Z>100$. The power law fit to fissility introduced by Swiatecki \cite{Swiatecki1955} (which also includes a shell correction term taken here to be the microscopic energy predicted by FRDM2012 \cite{FRDM2012}) also concentrates its high fission rates at $Z>100$. In contrast, the phenomenological description of \cite{XuRen}, which is parameterized as a function of fissility, $N-Z$, $N-126$, and $Z-82$, has dramatically different behavior, predicting a very large region of fast rates at $Z\gtrsim94$ just above the $N=184$ predicted shell closure. The bottom panel of Fig.  \ref{fig:spf} permits a comparison between these parameterized spontaneous fission fits and experimental data. Such parameterized descriptions clearly struggle to reproduce the rather abrupt order of magnitude variations seen in the experimental rates. While attempts to apply more physically meaningful approaches for the spontaneous fission rates of neutron-rich nuclei have been made in recent years \cite{Samuel}, ultimately the availability of theoretical data for this process remains scarce. We look forward to theoretical calculations and experimental measurements to better inform the treatment of spontaneous fission in the $r$ process in the future.

Since most descriptions show spontaneous fission to play its largest role at high proton number, consistent with other studies we find that $r$-process material tends to encounter spontaneous fission at late times when the abundances of almost all $r$-process nuclei are already finalized. However such late-time fission can have an impact on other $r$-process observables via the heat that a $\sim 200$ MeV Q-value will transfer to the astrophysical environment. For instance, \cite{Cfpaper} showed that the spontaneous fission of the long-lived species $^{254}$Cf has the potential to impact merger light curves. The light curve response to the spontaneous fission heating from $^{254}$Cf takes place on the order of 10 to 100 days and is presently the only way possible to confirm that a merger event produced elements heavier than lanthanides such as gold and uranium. However whether or not $^{254}$Cf can have such an impact depends on the fission barrier model assumed \cite{VasshJPG} since the nuclei on the $\beta$-decay path feeding this species can be depopulated from neutron-induced or $\beta$-delayed fission. Therefore, advancements in the theoretical description for the fission rates of neutron-rich isotopes is of great importance in understanding multi-messenger signals. At the present it is unclear whether other nuclei accessible to the $r$ process with spontaneous fission half-lives on the order of days exist in neutron-rich regions.

\section{Impact of fission yields on $r$-process abundances}\label{sec-3}

In very neutron-rich conditions, fission fragment distributions play a crucial role in determining the abundances of lighter elements near the $A\sim130$ second $r$-process peak (such as tellurium ($Z=52$)). This is largely due to a significant abundance of species present in the fissioning regions after the fast neutron capture of lighter elements pushes out to very high mass numbers just before the neutron flux is exhausted. These high mass species then deposit lower mass number daughter products into the main $r$-process pattern (between the second and third ($A\sim195$) $r$-process peaks) when they fission. In our calculations we consider fission yield daughter products at the 0.01\% level and above. We have found that the order of magnitude threshold at which $r$-process abundances are sensitive to the fission yields of neutron-rich nuclei is at the 0.1\% level since calculations which cut off yields at the 1\% level produce different abundance results.

\begin{figure}[h]
\centering
\includegraphics[scale=0.375]{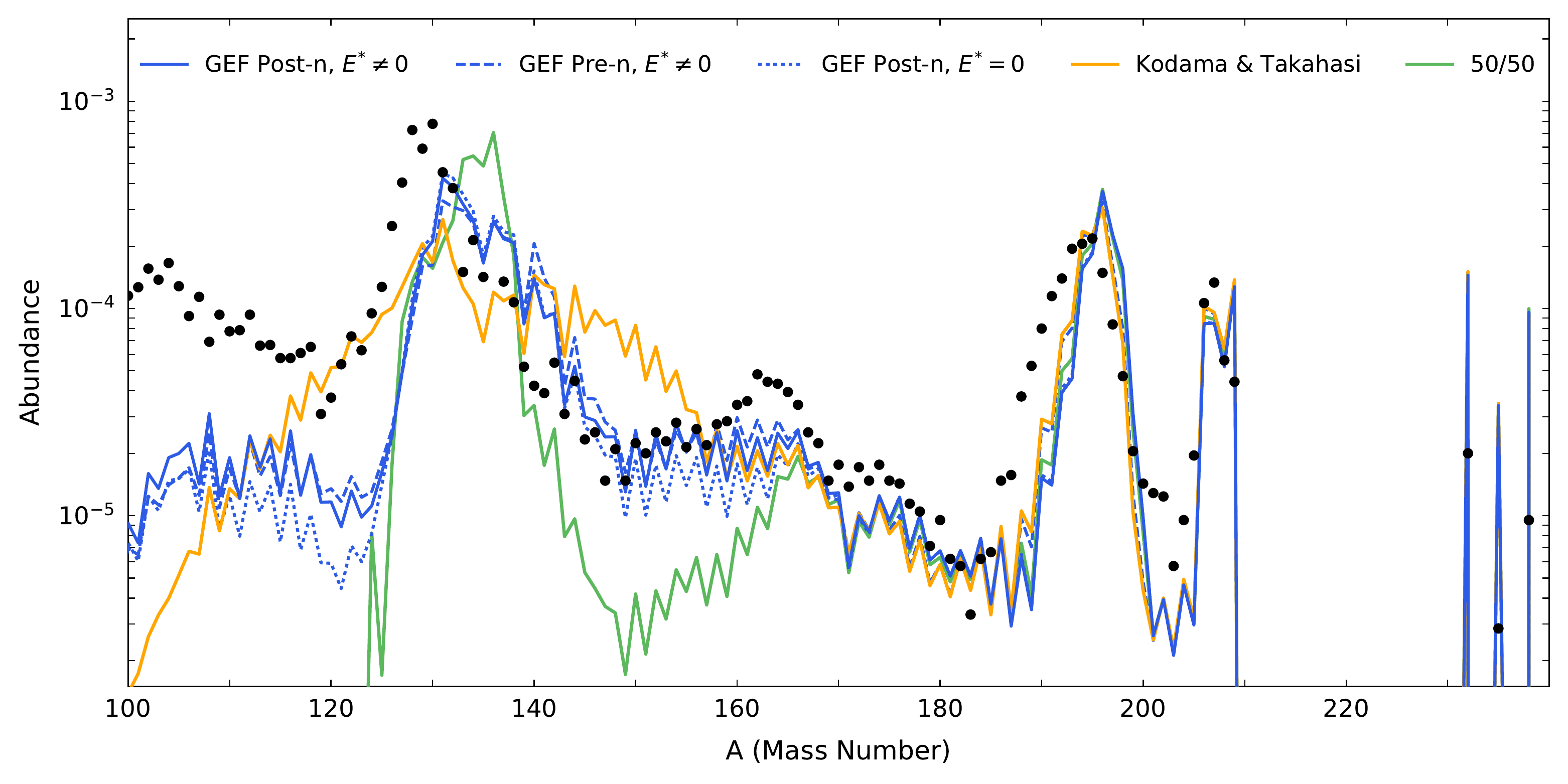}
\caption{Nucleosynthesis calculation abundances at 1 Gyr as a function of mass number for cold dynamical ejecta conditions as can be found in a merger tidal-tail \cite{Rosswog} using different prescriptions for the fission yields of neutron-rich nuclei: GEF2016 \cite{GEFnucdata} (blue), Kodama \& Takahashi \cite{KodTak} (yellow), and 50/50 symmetric yields where it is assumed that the nucleus simply splits in half (green). The dashed blue line examines the impact of ignoring prompt neutron emission by applying pre-neutron emission yields. The dotted blue line considers results when the excitation energy dependence of yields is ignored ($E^{*}=0$). Black dots are the solar abundance data from \cite{Sneden}.}
\label{fig:yieldcomp}
\end{figure}

As can be seen from the comparison in Fig. \ref{fig:yieldcomp}, the prescription for fission fragment distributions can lead to differences as large as an order of magnitude in the $r$-process abundances. In addition to the impact from the primary fragment yield treatment, we considered the sensitivity of $r$-process abundances to the excitation energy dependence of the fission yields for each distinct fission channel (neutron-induced, $\beta$-delayed, and spontaneous) in \cite{VasshJPG} using the GEF2016 code \cite{GEFnucdata}. Prior to this work, common practice for $r$-process calculations was to apply one yield set to all channels and neglect the excitation-energy dependence of fission yields. Additionally we considered the impact of prompt neutron emission via direct comparisons of results with GEF and FREYA \cite{FREYA,RamonaProc} for neutron-rich nuclei. As shown in Fig.~\ref{fig:yieldcomp}, we found that considering yields which have a non-vanishing excitation energy was of greater impact on $r$-process abundances than the de-excitation treatment which determines prompt neutron emission.

\section{Fission yields from macroscopic-microscopic theory in the $r$ process}\label{sec-4}

\begin{figure}[h]
\centering
\includegraphics[scale=0.32]{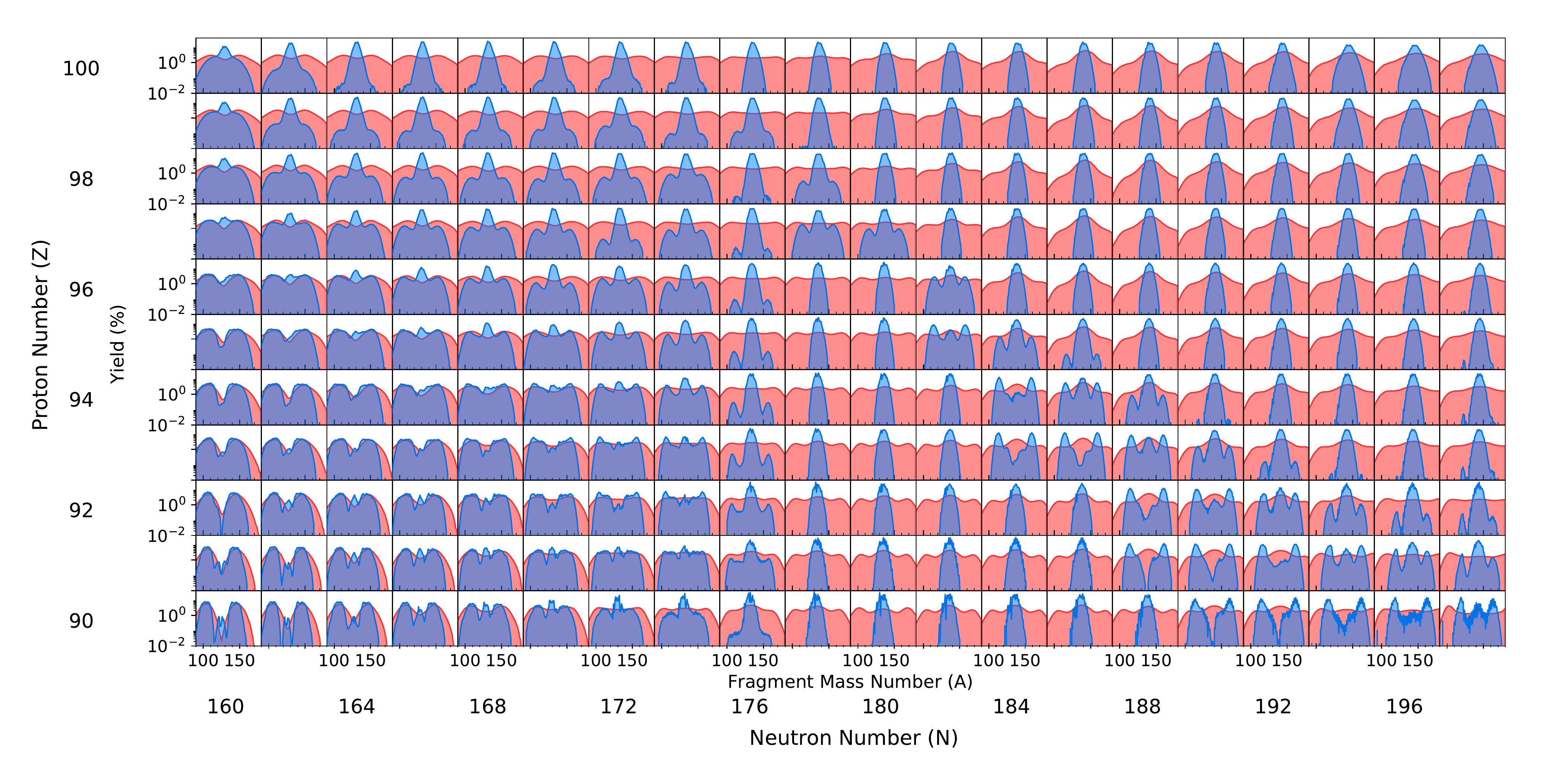}
\caption{The fission yields predicted for some neutron-rich nuclei of relevance in the $r$ process given recent results which make use of the macroscopic-microscopic theory of FRLDM \cite{MattFRLDM} (red) as compared to the data-driven, phenomenological calculations of GEF2016 (neutron-induced fission with an incident neutron energy of 0.1 MeV) \cite{GEFnucdata} (blue).}
\label{fig:frldmgef}
\end{figure}

Very few theoretical approaches predicting fission yields have been applied to neutron-rich nuclei \cite{Goriely,MattFRLDM,MicroArxiv} and only a subset have considered all of the neutron-rich isotopes needed to perform a nucleosynthesis calculation. Therefore the majority of calculations in the literature make use of data-driven phenomenological models, such as the GEF treatment presented in the last section as well as the ABLA code (the precursor of GEF) and the Wahl systematics (see \cite{VasshJPG} for references and a literature overview of fission yield applications in the $r$ process). Recently, fission yields calculated using the FRLDM macroscopic-microscopic approach \cite{FRLDM} have been reported for the broad range of neutron-rich isotopes of relevance in the $r$ process \cite{MattFRLDM}. These fragment distributions have a wider range of daughter products than previously predicted by phenomenological extrapolations such as GEF2016, as can be seen in Figure~\ref{fig:frldmgef}.

\begin{figure}[h]
\centering
     \includegraphics[width=11cm]{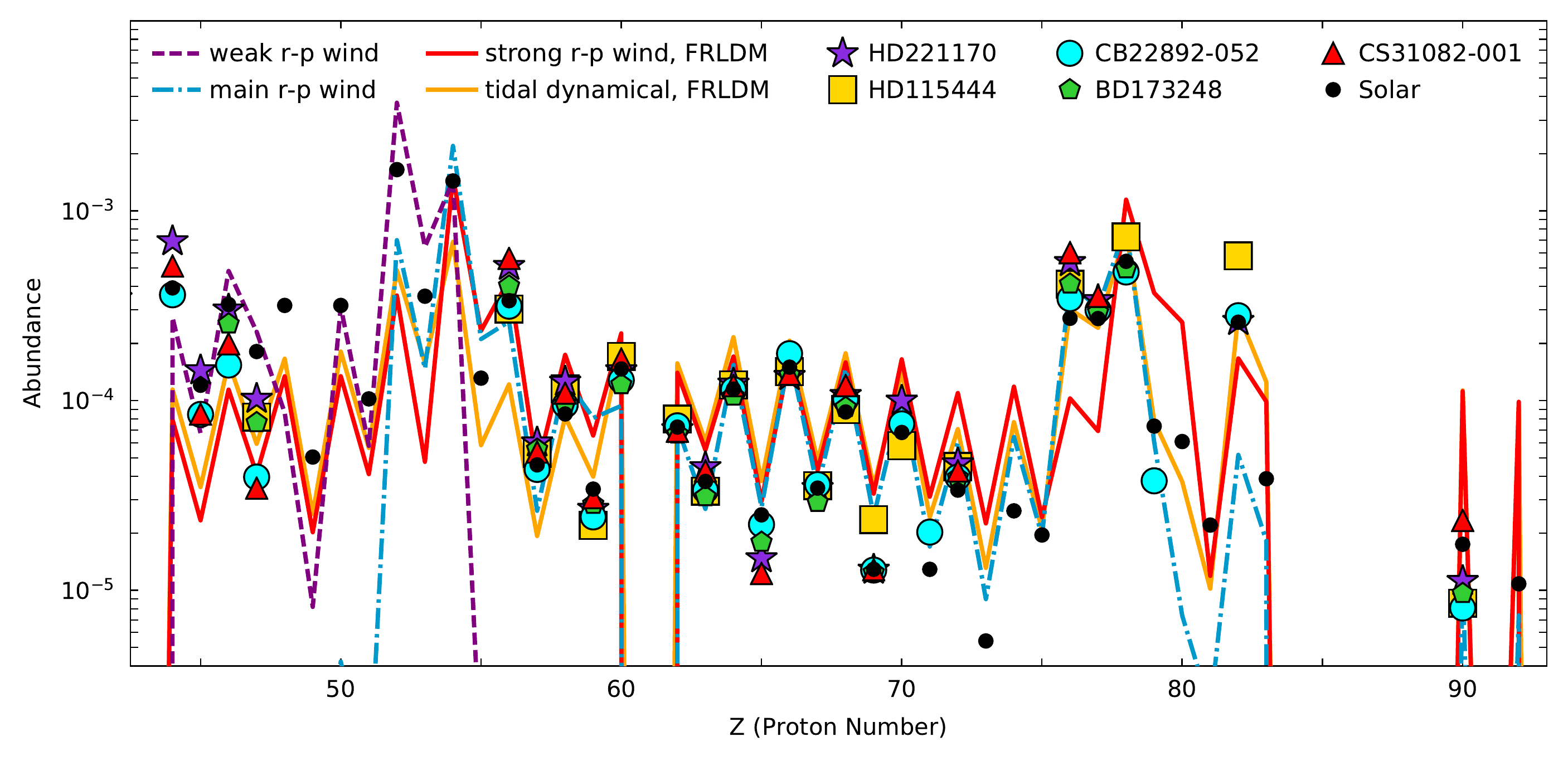} %
     \includegraphics[width=11cm]{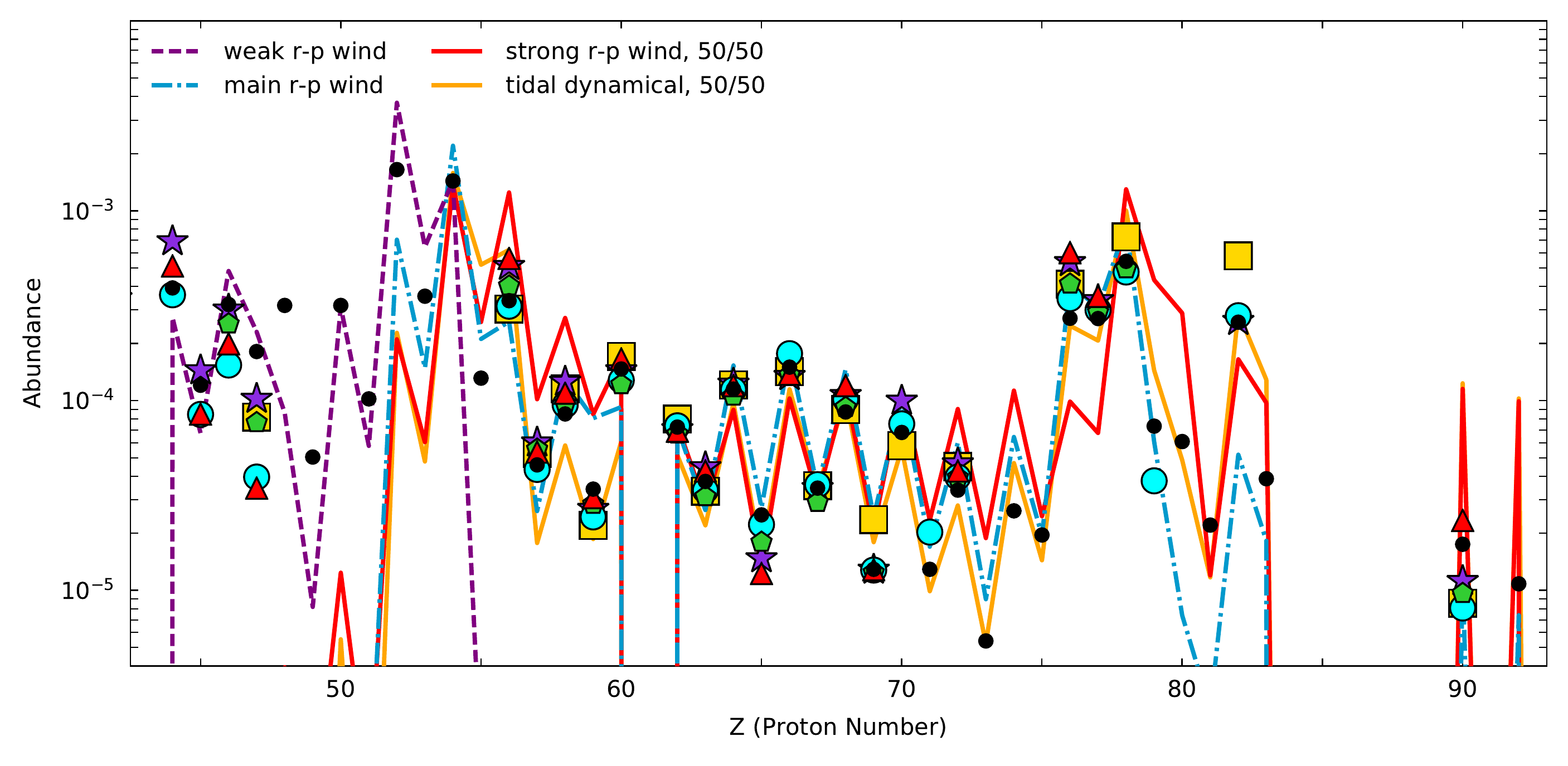}%
 \caption{The elemental abundances predicted by nucleosynthesis calculations at 1 Gyr for dynamical merger tidal-tail ejecta \cite{Rosswog} (orange) as compared to three distinct conditions all present within a simulation of accretion disk wind ejecta \cite{JustWind}: a weak $r$ process which does not synthesize elements beyond the second $r$-process peak (purple dashed), a main $r$ process which mostly populates nuclei between the second and third $r$-process peaks and weakly produces actinides (blue dashed-dotted) and a strong $r$ process which robustly reaches very neutron-rich, fissioning species. The top panel shows results using FRLDM fission yields and the bottom panel shows these cases using 50/50 symmetric splits. The solar and stellar abundance data is from \cite{Sneden}.}
\label{fig:abZfrldm5050}
\end{figure}

The elemental $r$-process abundances using FRLDM fission yields, as compared to applying a simple 50/50 symmetric split for fissioning nuclei, are shown in Fig. \ref{fig:abZfrldm5050} for several distinct astrophysical conditions. A merger tidal tail is often found by simulation to be very neutron-rich and therefore can easily synthesize up to the fissioning isotopes. Calculations for such a condition demonstrate the ability of the wide yields of FRLDM to not only populate the main $r$-process nuclei between the second and third peaks, but also significantly produce the light precious metals such as palladium ($Z=46$) and silver ($Z=47$). Such co-production of light precious metals and lanthanide elements ($Z=57$ to $Z=71$) is not observed for more simple fission yield prescriptions such as 50/50 splits. Also shown are abundances given three distinct accretion disk wind trajectories, all of which can be found in the simulation of \cite{JustWind}. In this wind simulation, most trajectories produce abundances similar to the `weak' $r$-process case which does not populate nuclei beyond the second $r$-process peak. It is clear from the case examining 50/50 yield results that without contributions to the light precious metals from fission deposition the only way to populate such species at a given astrophysical site is for some of the ejecta to undergo solely a weak $r$ process. Given naturally occurring variations at astrophysical sites such as the mass of the neutron stars and the nature of the merger remnant, the amount of weak $r$-process ejecta relative to ejecta which produces heavier species such as the lanthanides and actinides can also naturally vary. However if conditions which access fissioning nuclei are significantly present at a site, the co-production of weak $r$-process elements and lanthanides gives final abundances which are more immune to variances in weak versus main $r$-process ejecta. This stabilizing effect on the final abundances could help explain the so-called `universality' of elemental abundances observed in $r$-process enhanced, metal-poor stars \cite{VasshFRLDMrp}.

\section{Possible signature of co-production via fission in $r$-process enhanced, metal-poor stars}\label{sec-5}

Observations of the elemental abundances in our Sun and other stars can provide insights as to the nature of the site(s) producing heavy elements. Examining metal-poor stars which are rich in $r$-process material (i.e. r-I and r-II stars \cite{JINAbase}) reveals the elemental production from one to few nucleosynthesis events since such stars have not yet been significantly polluted by events such as supernovae which inject metal-rich material into the interstellar medium. Elemental patterns for such stars show a remarkable consistency with each other and the content of $r$-process material in our Sun from barium ($Z=56$) through the third $r$-process peak \cite{Sneden}. This `robustness' or `universality' has been discussed as a potential signature of fission in the $r$ process (see \cite{VasshFRLDMrp} and references therein) since fission deposition back into lighter elements near $A\sim130$ can potentially wash away differences in the initial conditions prior to the synthesis of fissioning species. 

In considering a possible connection between fission and universality, it is important to identify what feature of stellar abundances universality statements are in fact referencing. For instance, if universality refers to similarities in the r-I and r-II stellar data for the abundances of $r$-process nuclei, fission may not be the process producing this since in \cite{VasshJPG} we demonstrated that astrophysical conditions which robustly reach fissioning nuclei can have varying abundance patterns. This is in part due to distinct conditions populating different fissioning species (implying distinct fission yield distributions set abundances when depositing daughter products). Differences in conditions also determine how much of the second peak is formed by late-time fission as well as how much late-time neutron capture can shift the second peak. It is unclear how large such differences in abundances can be in an actual astrophysical environment hosting fission given the variance seen in the conditions reported by astrophysical simulations as well as the uncertainties in the fission barriers and fission yields of neutron-rich nuclei. Additionally it is unclear whether the uncertainties in stellar observations of elemental abundances overshadow the abundance differences referred to here from varying astrophysical conditions.

If instead universality statements are in reference to the abundances of lanthanides as compared to abundances of elements in the third $r$-process peak, fission can still be a means to explain the consistency in the stellar data since all fissioning conditions robustly produce the full range of $r$-process nuclei. However, some universality arguments solely consider lanthanide abundances relative to each other since elemental patterns here are indeed remarkably similar from star to star. In fact, some authors have made use of stellar data to determine the solar $r$-process element content for lanthanides such as hafnium ($Z=72$) and gadolinium ($Z=64$) \cite{Sneden}. Here we examine whether universality in the lanthanides alone can be linked to fission. We also consider whether comparisons between the light precious metals to the left of the second $r$-process peak and the lanthanides can shed light on the prospect of late-time fission deposition in the $r$ process.

\begin{figure}[h]
\centering
     \includegraphics[width=6.1cm]{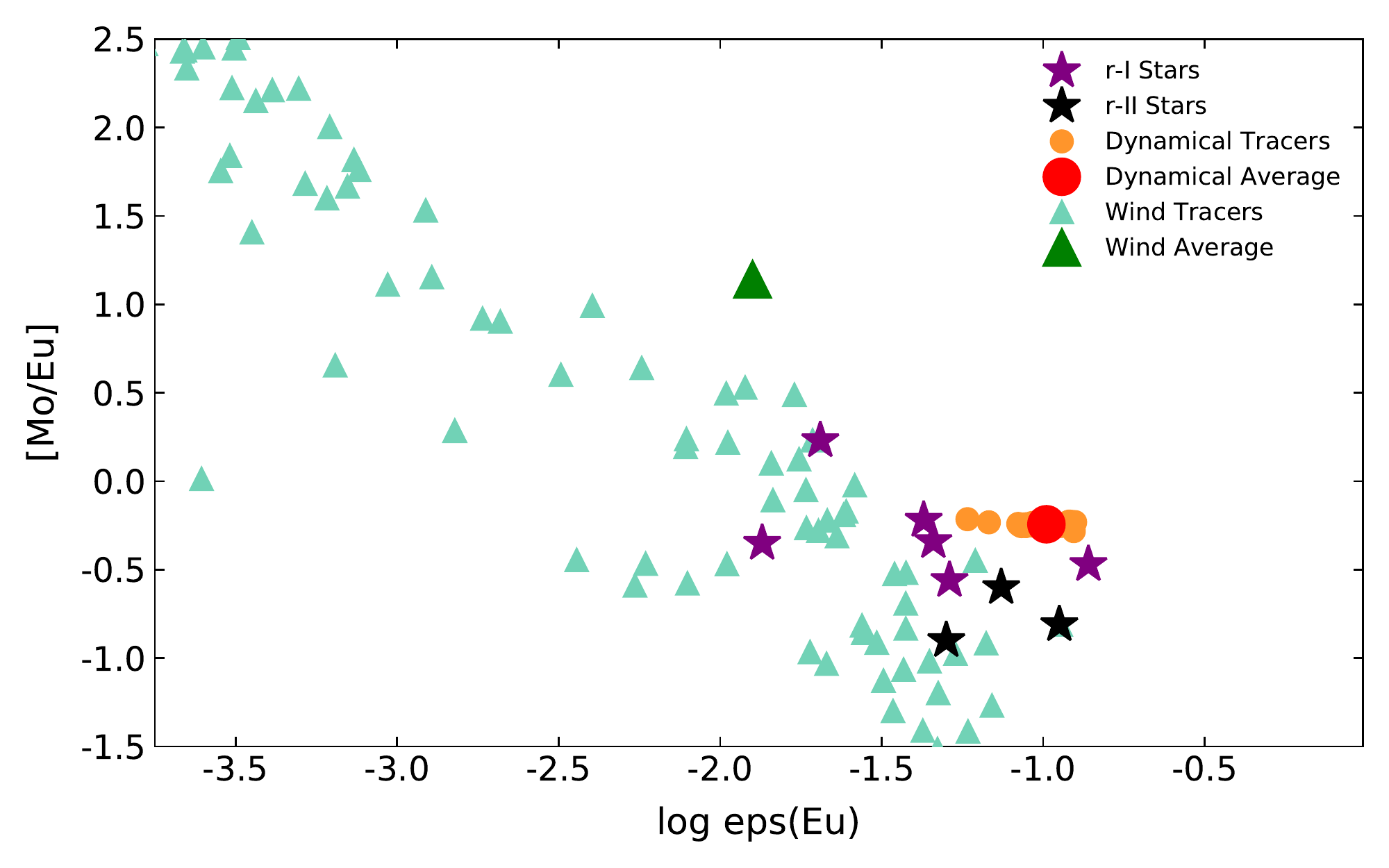} %
     \includegraphics[width=6.1cm]{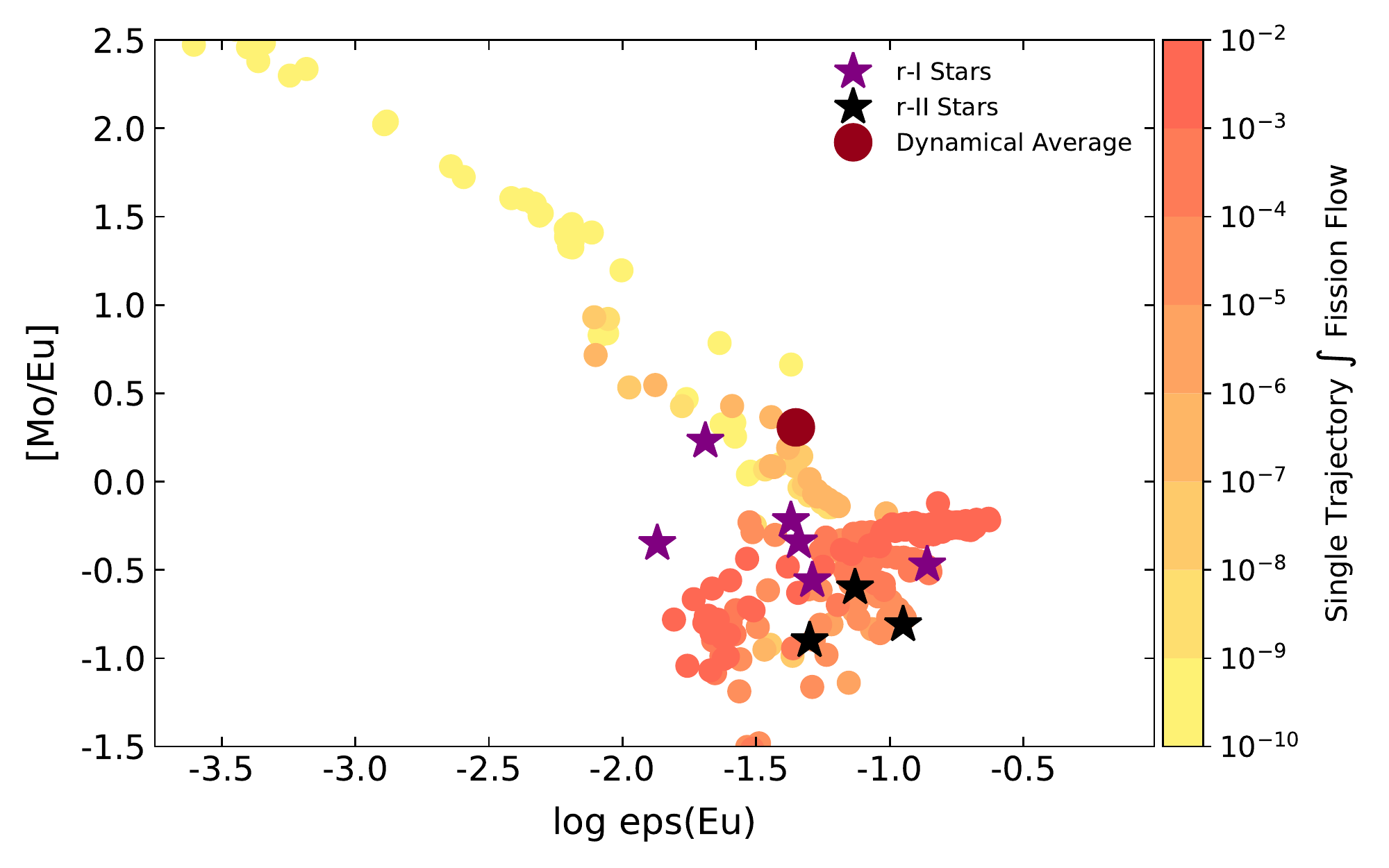}%
    \hspace{0.5cm}
     \includegraphics[width=6.1cm]{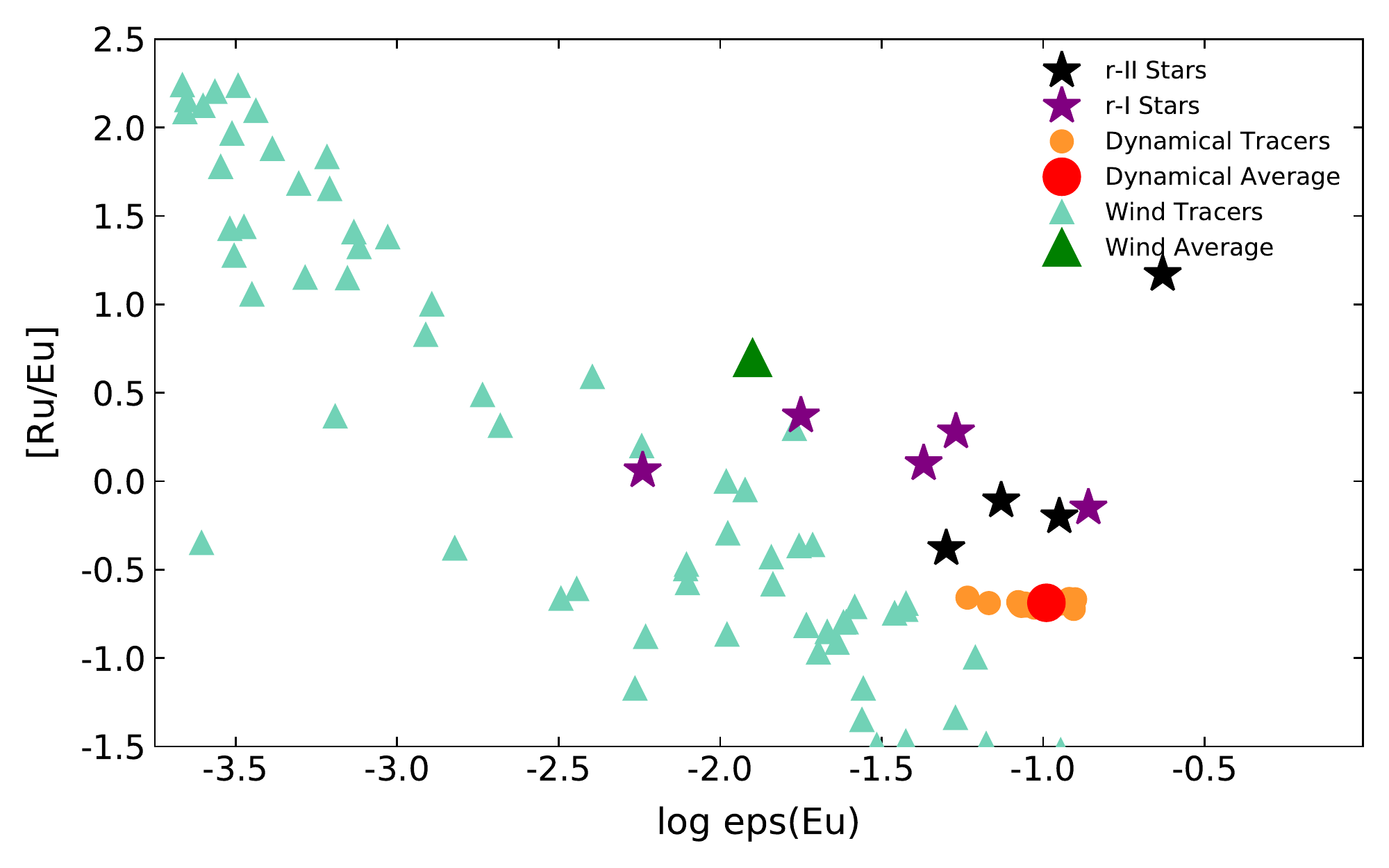} %
     \includegraphics[width=6.1cm]{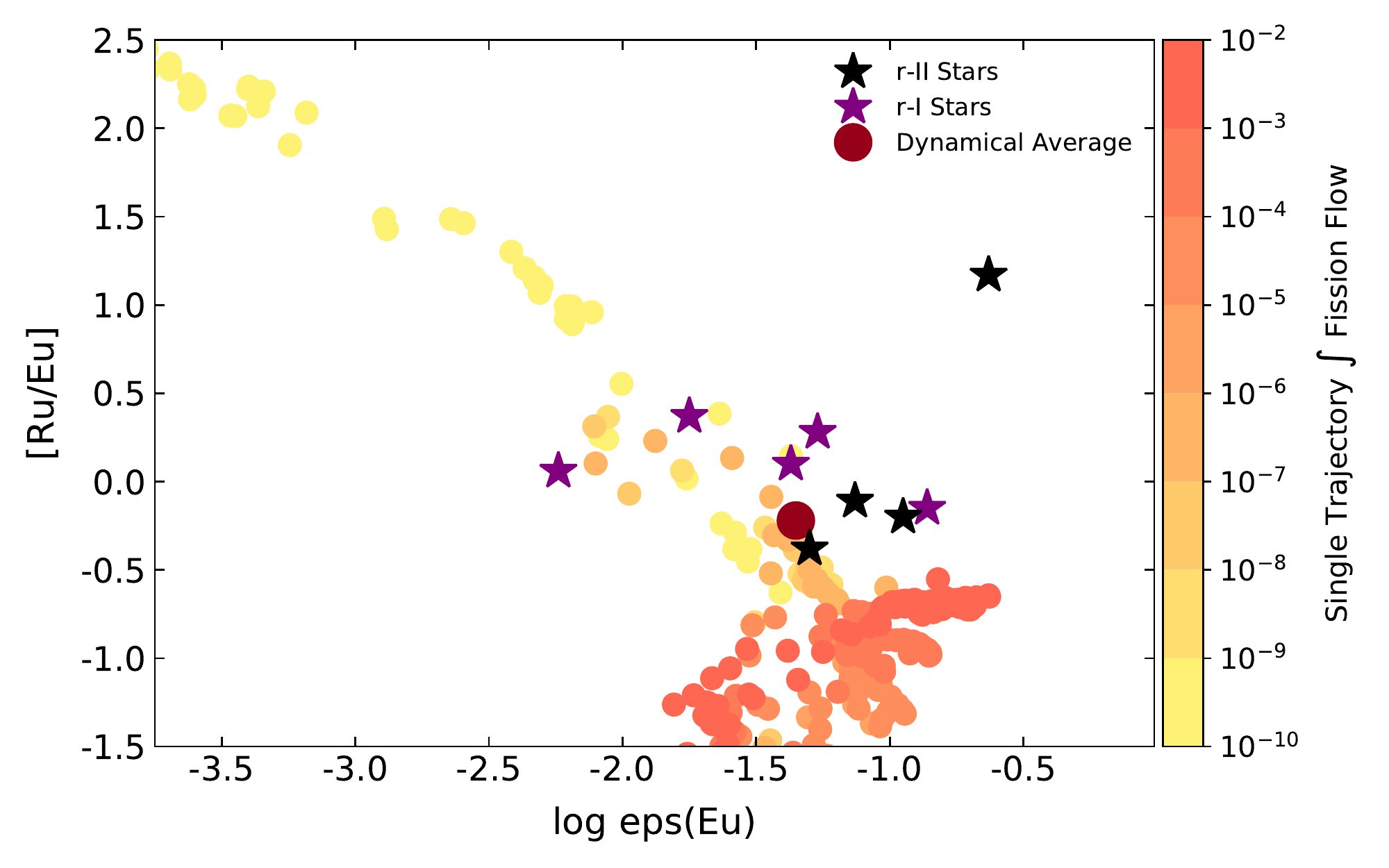}%
    \hspace{0.5cm}
     \includegraphics[width=6.1cm]{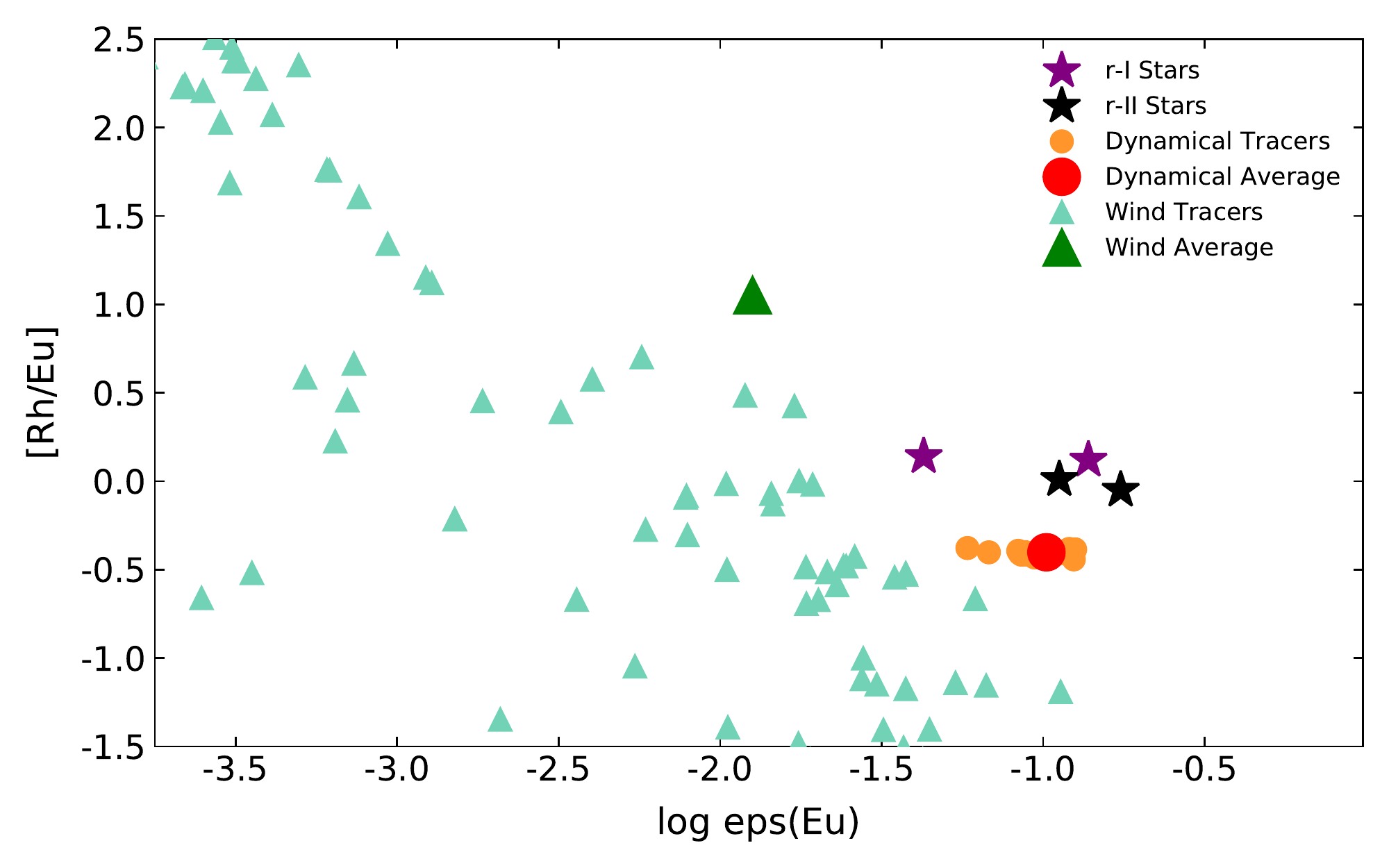} %
     \includegraphics[width=6.1cm]{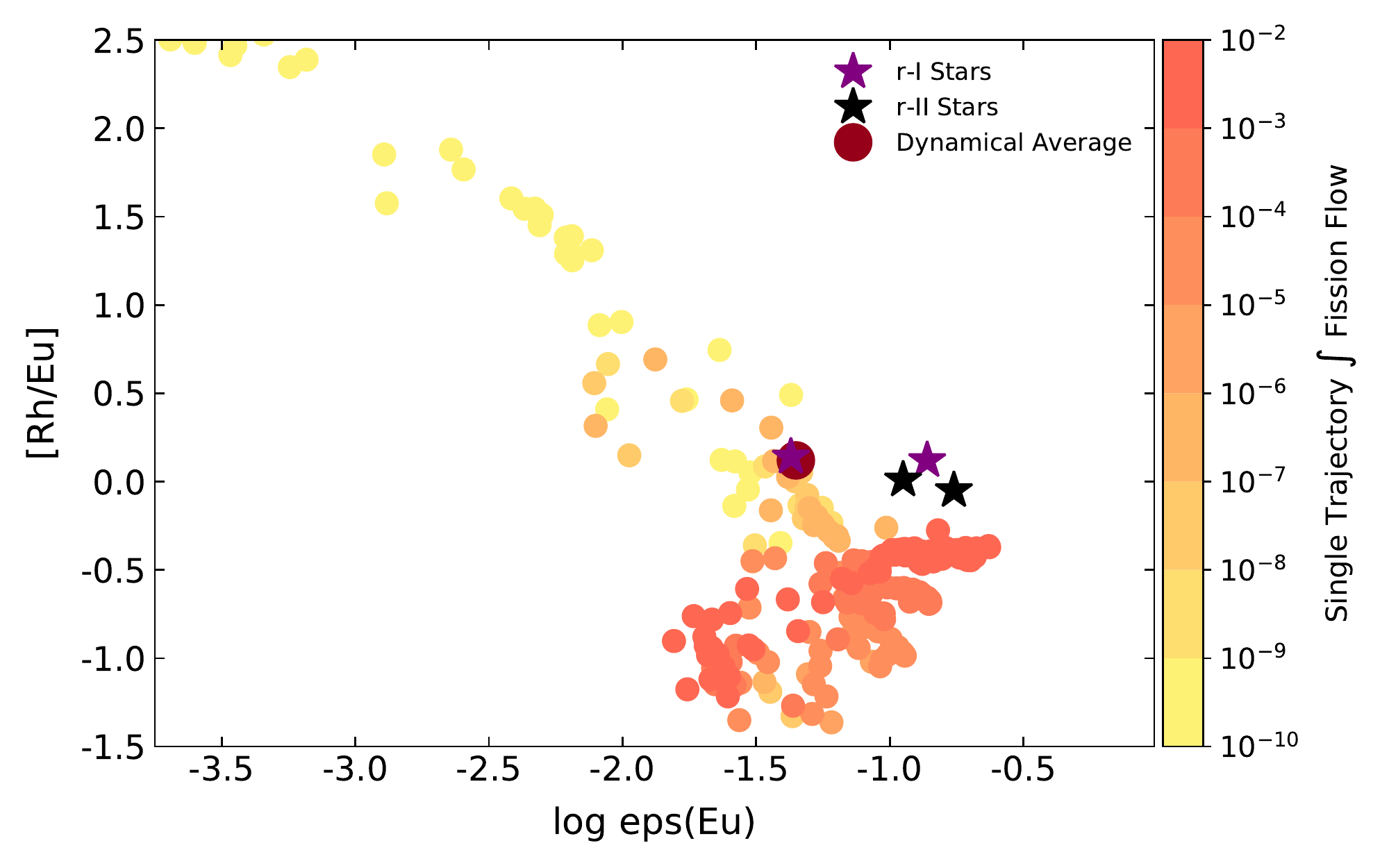}%
    \hspace{0.5cm}
     \includegraphics[width=6.1cm]{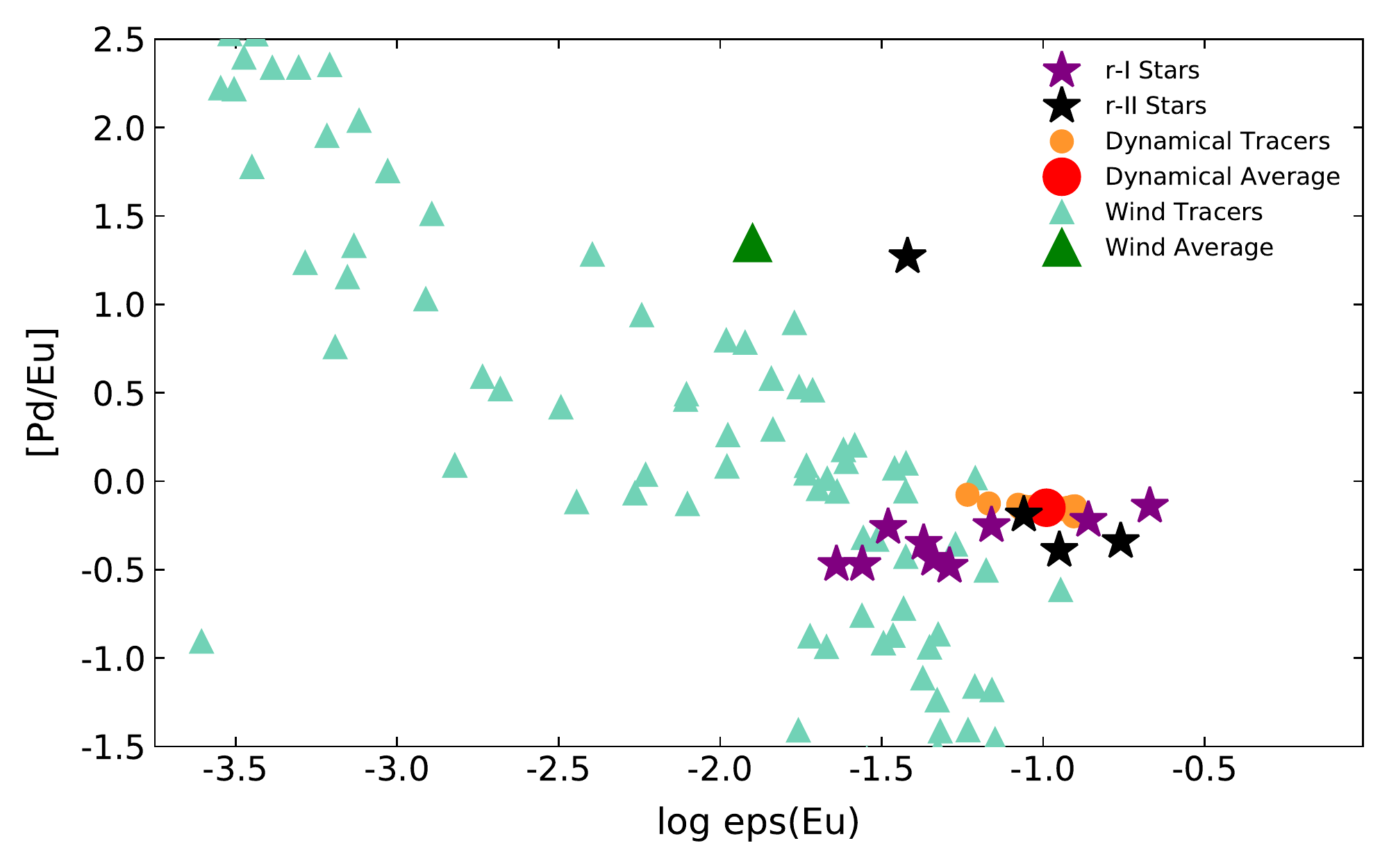} %
     \includegraphics[width=6.1cm]{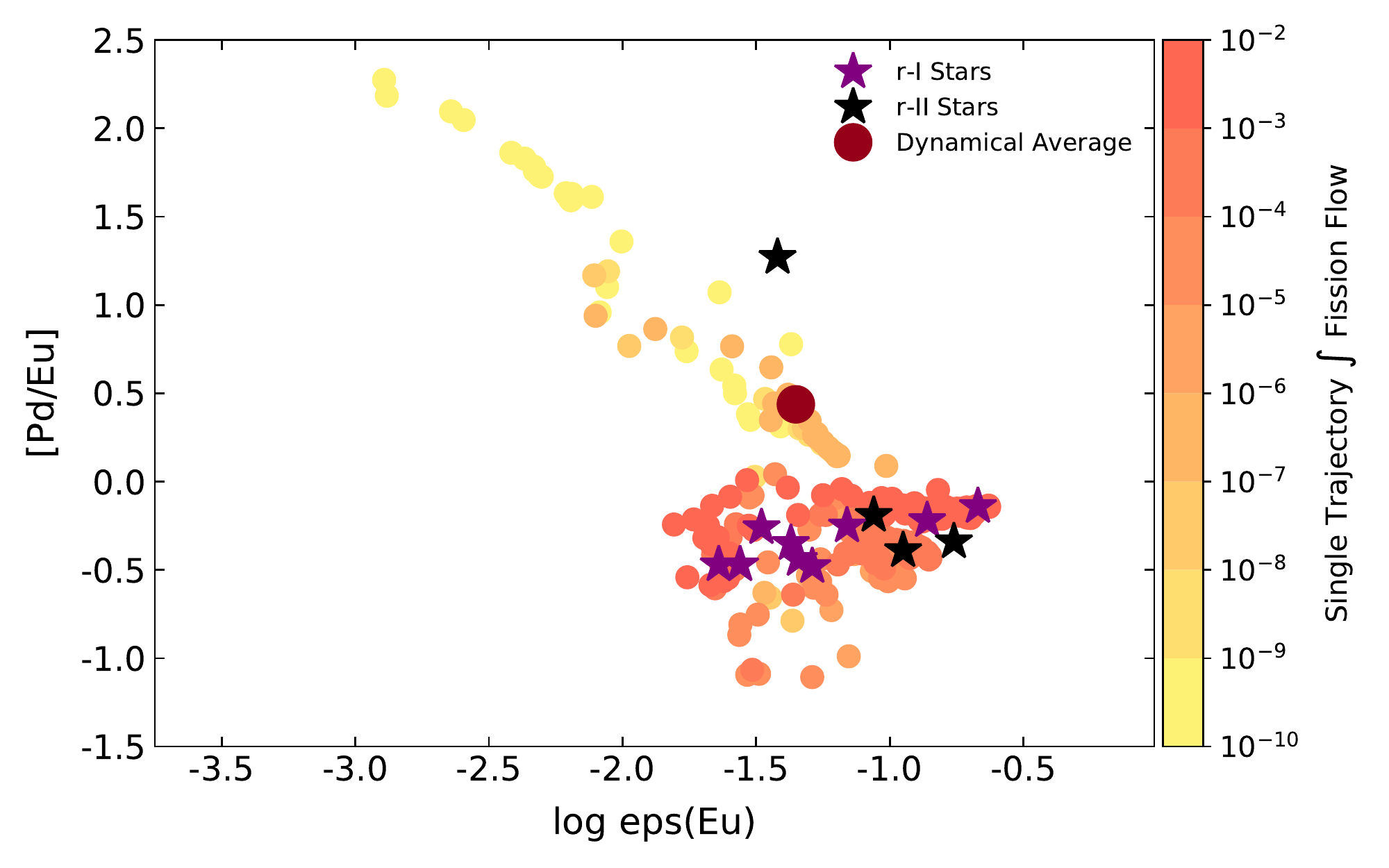}%
    \hspace{0.5cm}
     \includegraphics[width=6.1cm]{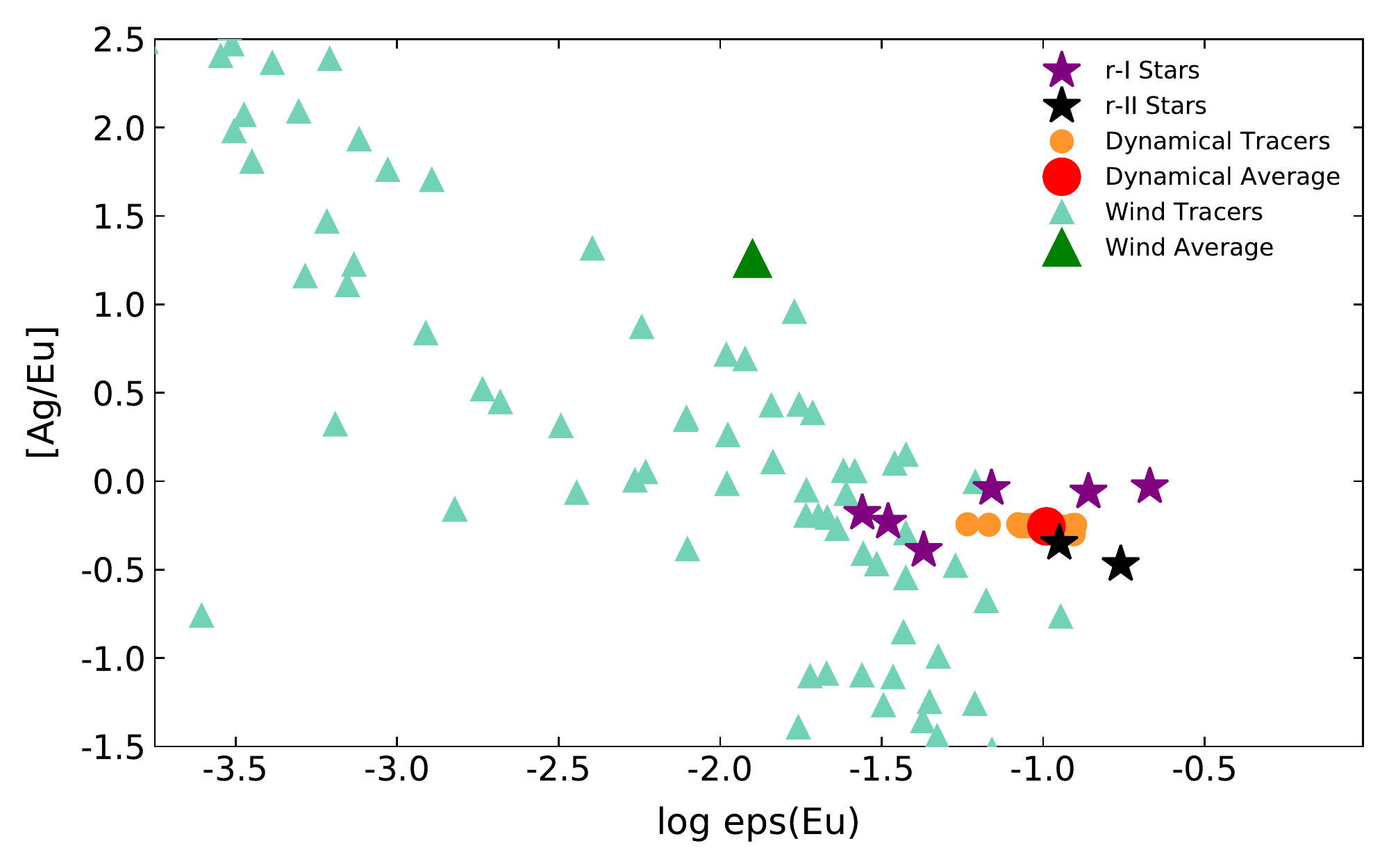} %
     \includegraphics[width=6.1cm]{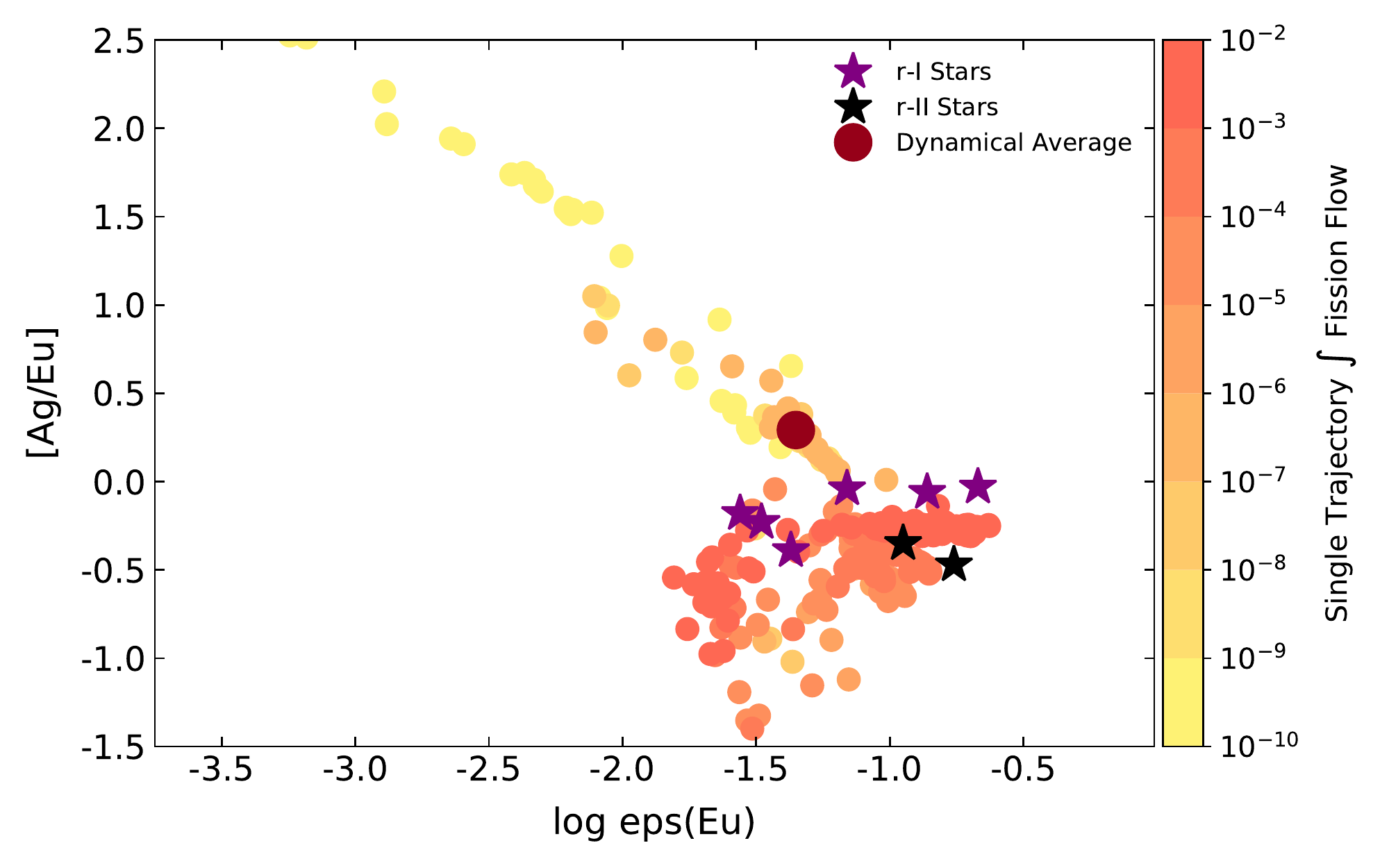}%
 \caption{Calculation results for the elemental ratios of light heavy elements (Mo, Ru, Rh, Pd, and Ag) found to the left of the second $r$-process peak relative to the lanthanide element Eu found to the right of the peak (a subset of such plots was shown in \cite{VasshFRLDMrp}). Observational values for stars which are metal-poor, yet rich in $r$-process elements are shown for comparison. See text for details.}
\label{fig:firstrIrII}
\end{figure}

We show the stellar data \cite{JINAbase} for the elemental ratios for some light heavy elements (Mo ($Z=42$), Ru ($Z=44$), Rh ($Z=45$), Pd ($Z=46$), and Ag ($Z=47$)) as compared to the lanthanide element, Eu ($Z=63$) in Fig.~\ref{fig:firstrIrII}. We compare to the ratio predicted by nucleosynthesis calculations. Note in these figures we use log eps(Eu) (the europium abundance) to quantify the $r$-process enrichment and the stellar values are that reported by observation. The values shown for the nucleosynthesis calculations must be scaled according to the absolute amount of europium, which depends on the amount of ejected matter from the astrophysical scenario considered, and this is presently unknown. Therefore in considering the agreement between stellar data and calculated values, only the ratios on the y-axis can be compared in a physically meaningful way.

Left panels of Figure~\ref{fig:firstrIrII} use the wind calculations of \cite{JustWind} as an example of conditions which primarily produce a weak $r$ process and most significantly populate nuclei lighter than those in the second $r$-process peak. Also shown are results with dynamical ejecta from \cite{Rosswog} for which all conditions robustly reach fissioning nuclei. Right panels consider the nucleosynthesis productions given a more modern dynamical ejecta simulation from \cite{Radice} which reports a wider range of conditions thereby producing both a weak and full $r$ process. Right panel nucleosynthesis results are color-coded by the total integrated fission flow (rate times abundance) in order to highlight the conditions present which most strongly access fissioning nuclei. The spread in the observational data for elements such as Mo and Ru suggests these elements to been produced independently of Eu, possibly by sites which do not synthesize past the second $r$-process peak. For the heaviest of the light heavy elements shown here, Pd and Ag, the flat trend seen in the stellar data suggests that such elements are co-produced with Eu in the astrophysical events which enriched these stars. Our nucleosynthesis calculations see conditions which are most consistent with observed ratios to be those which co-produce light precious metals and lanthanides via late-time fission deposition.

\begin{figure}[h]
\centering
     \includegraphics[width=6.1cm]{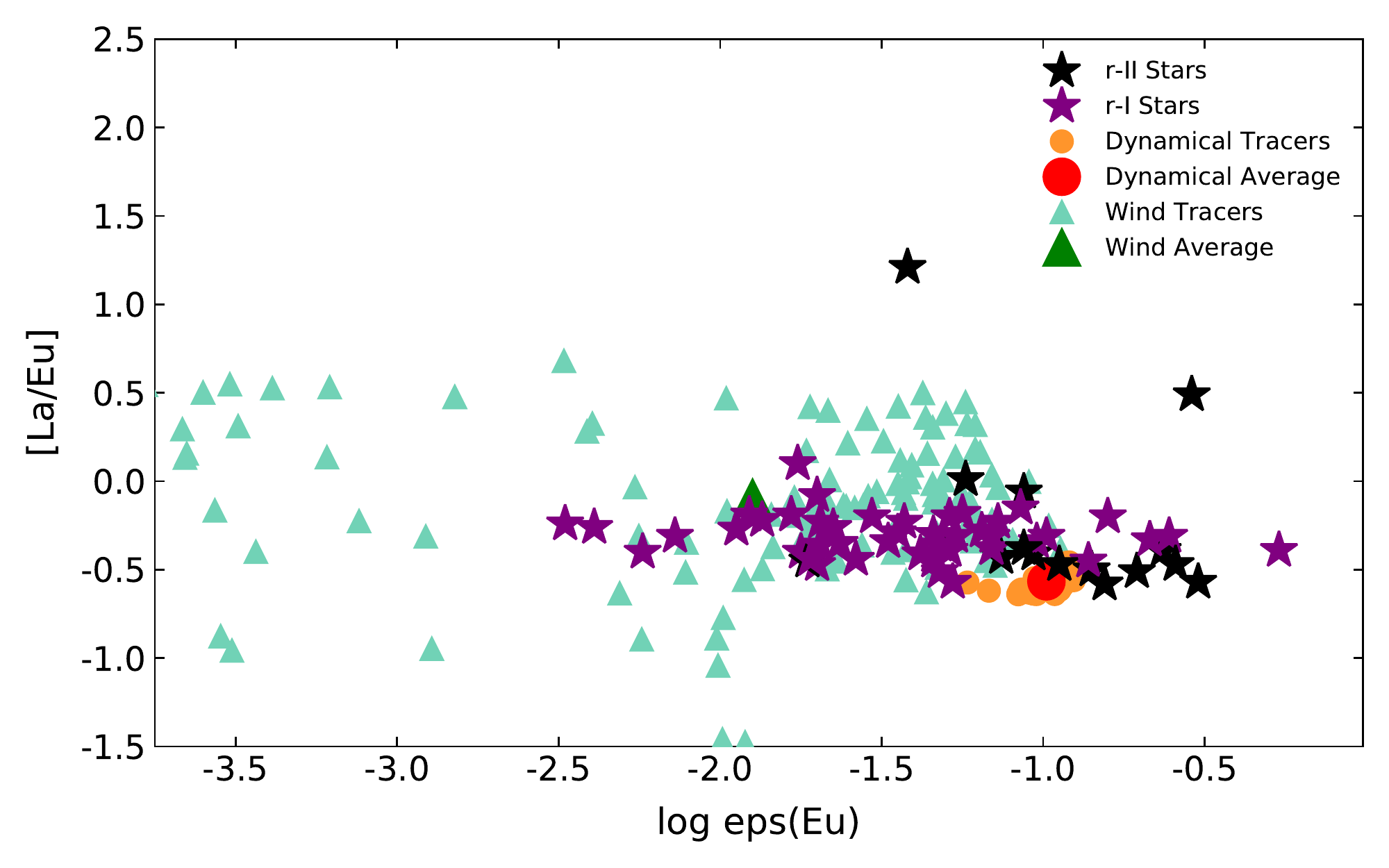} 
     \includegraphics[width=6.1cm]{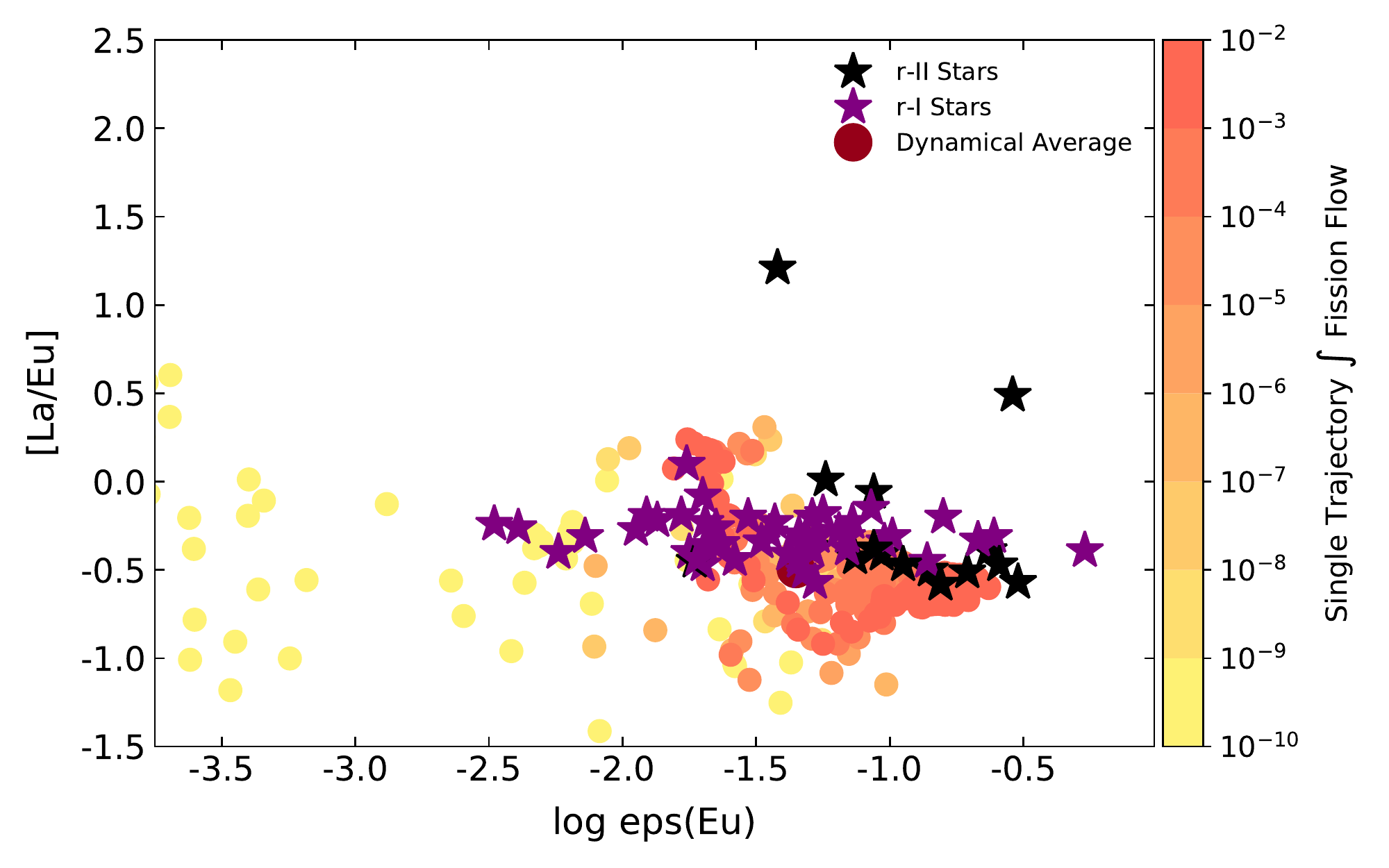}%
    \hspace{0.5cm}
    \includegraphics[width=6.1cm]{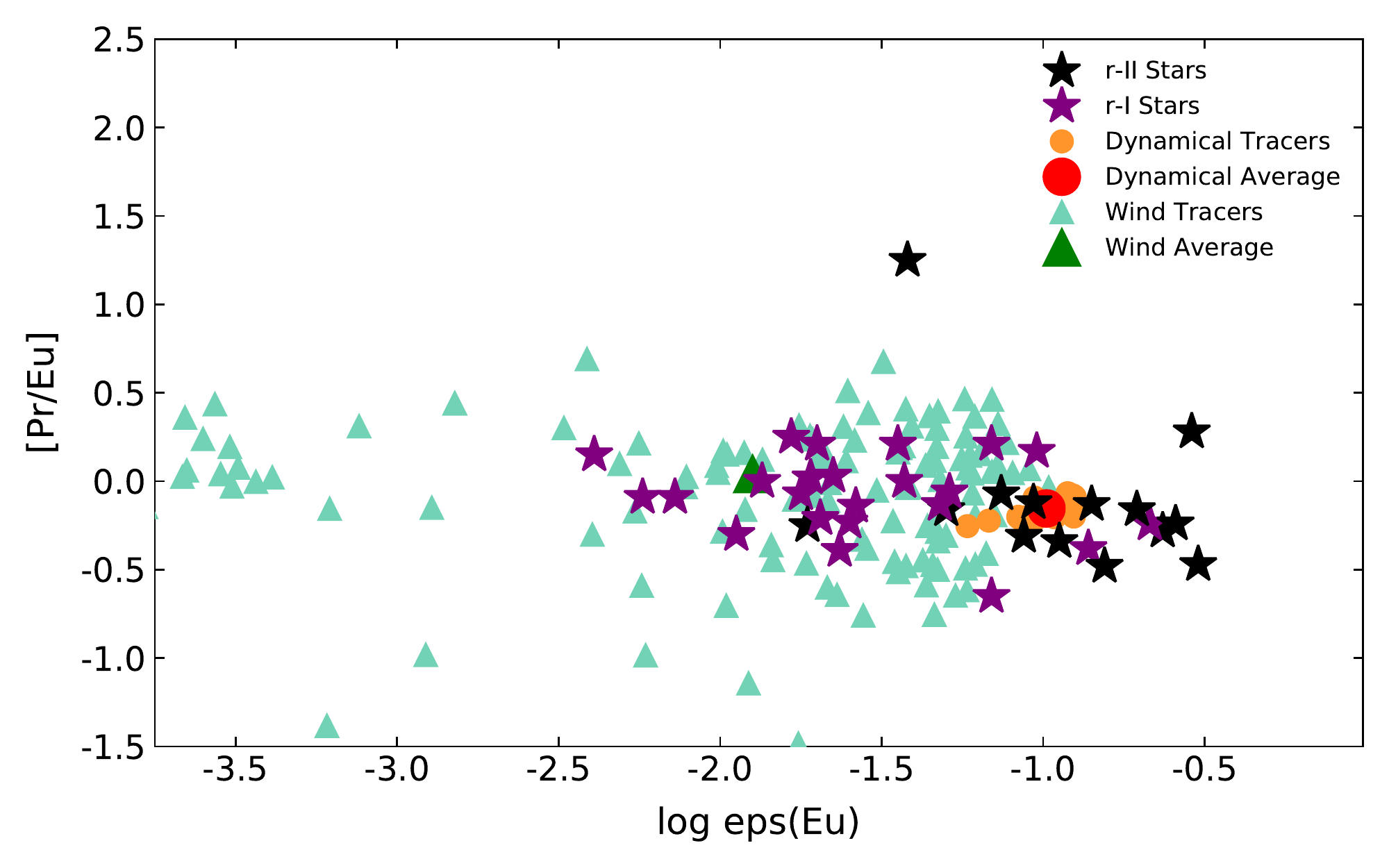} %
     \includegraphics[width=6.1cm]{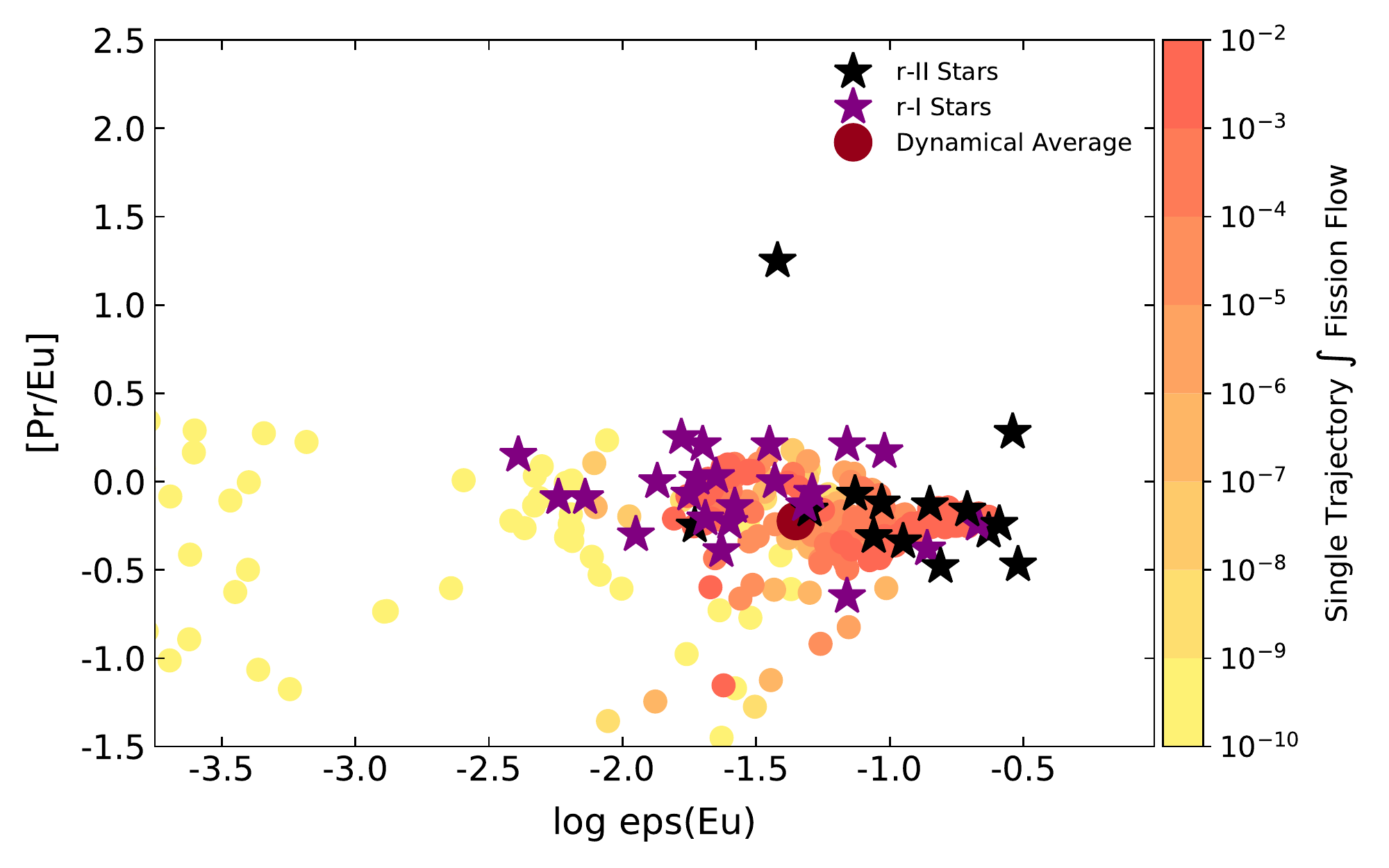}%
    \hspace{0.5cm}
      \includegraphics[width=6.1cm]{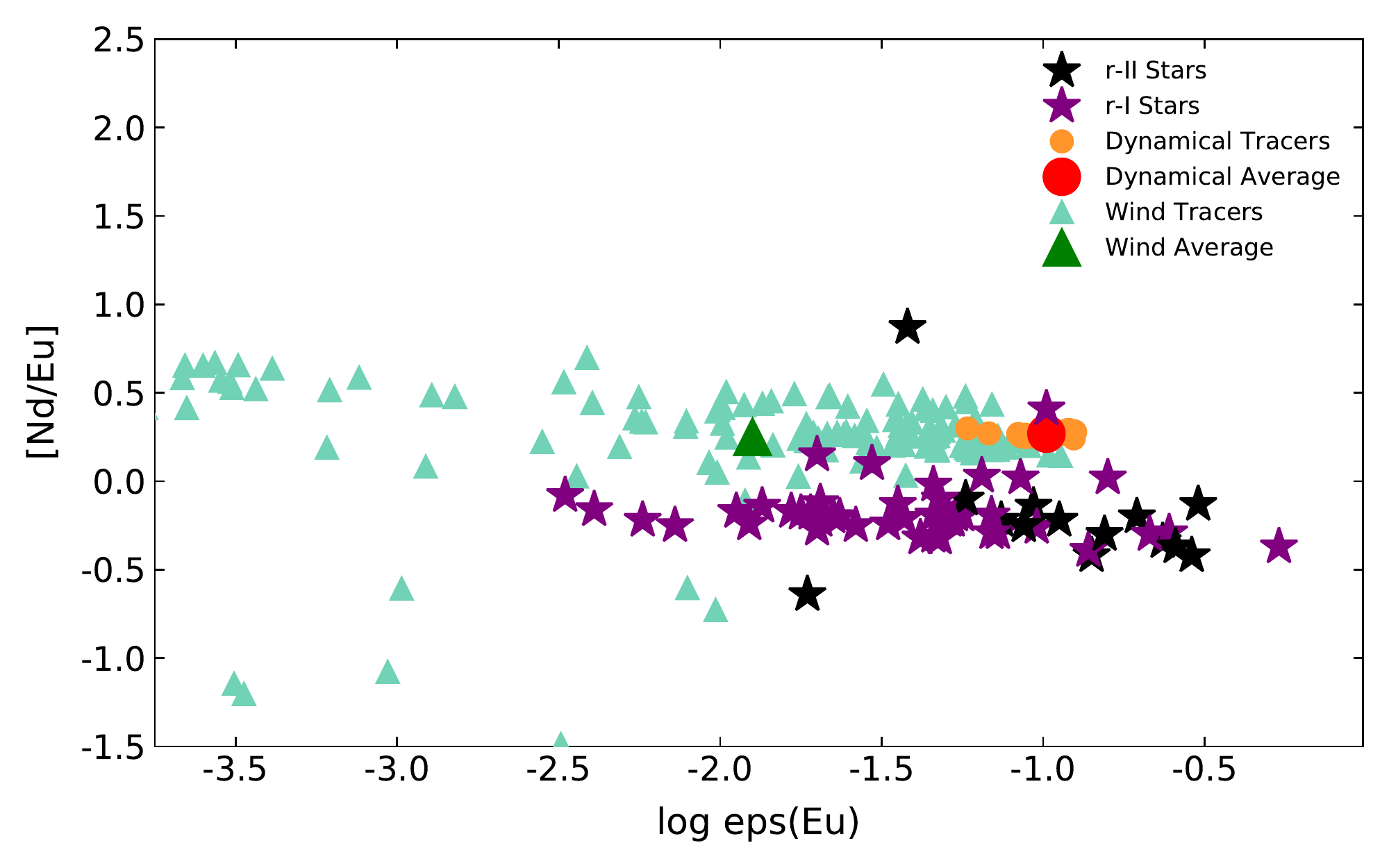} %
     \includegraphics[width=6.1cm]{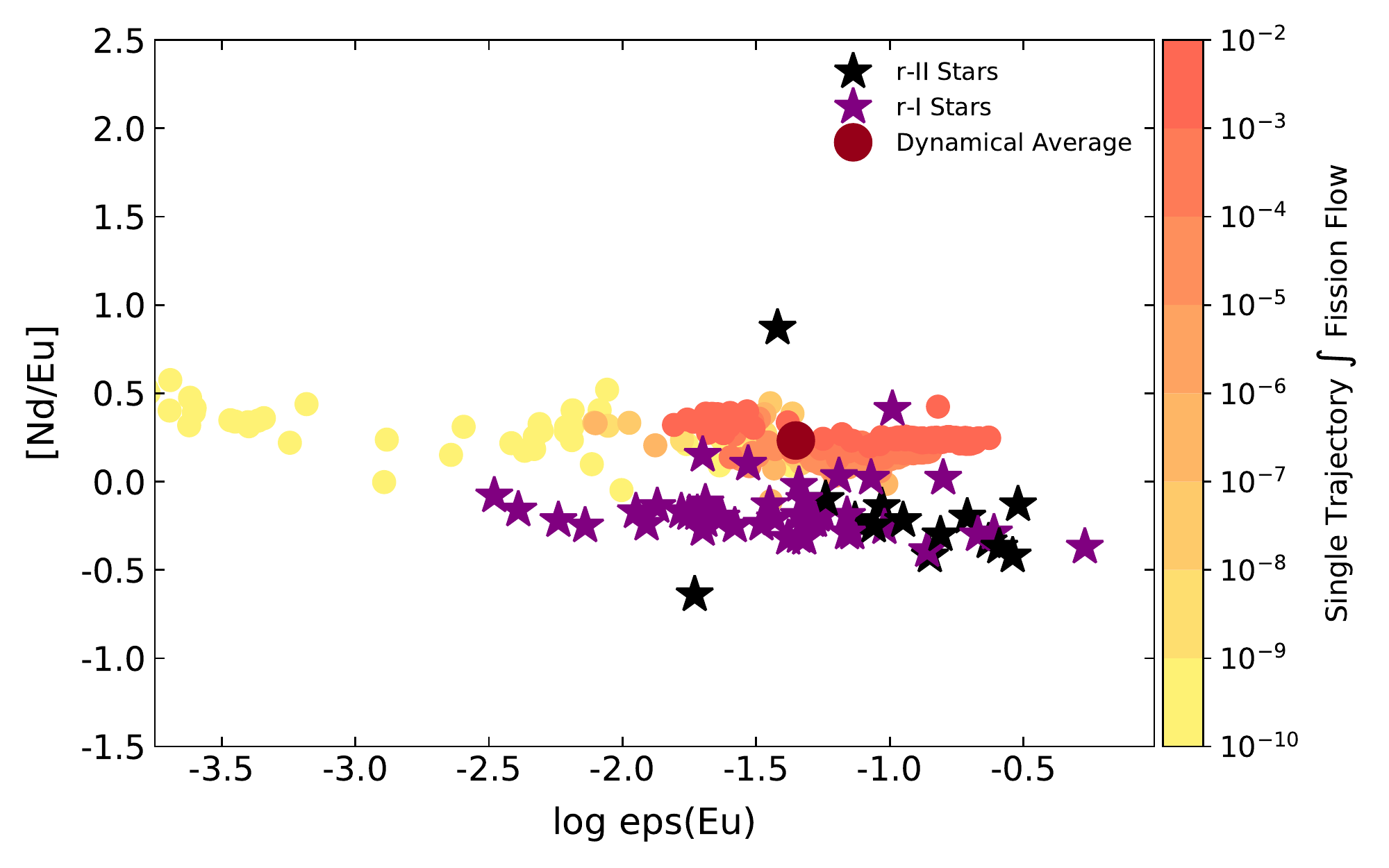}%
    \hspace{0.5cm}
    \includegraphics[width=6.1cm]{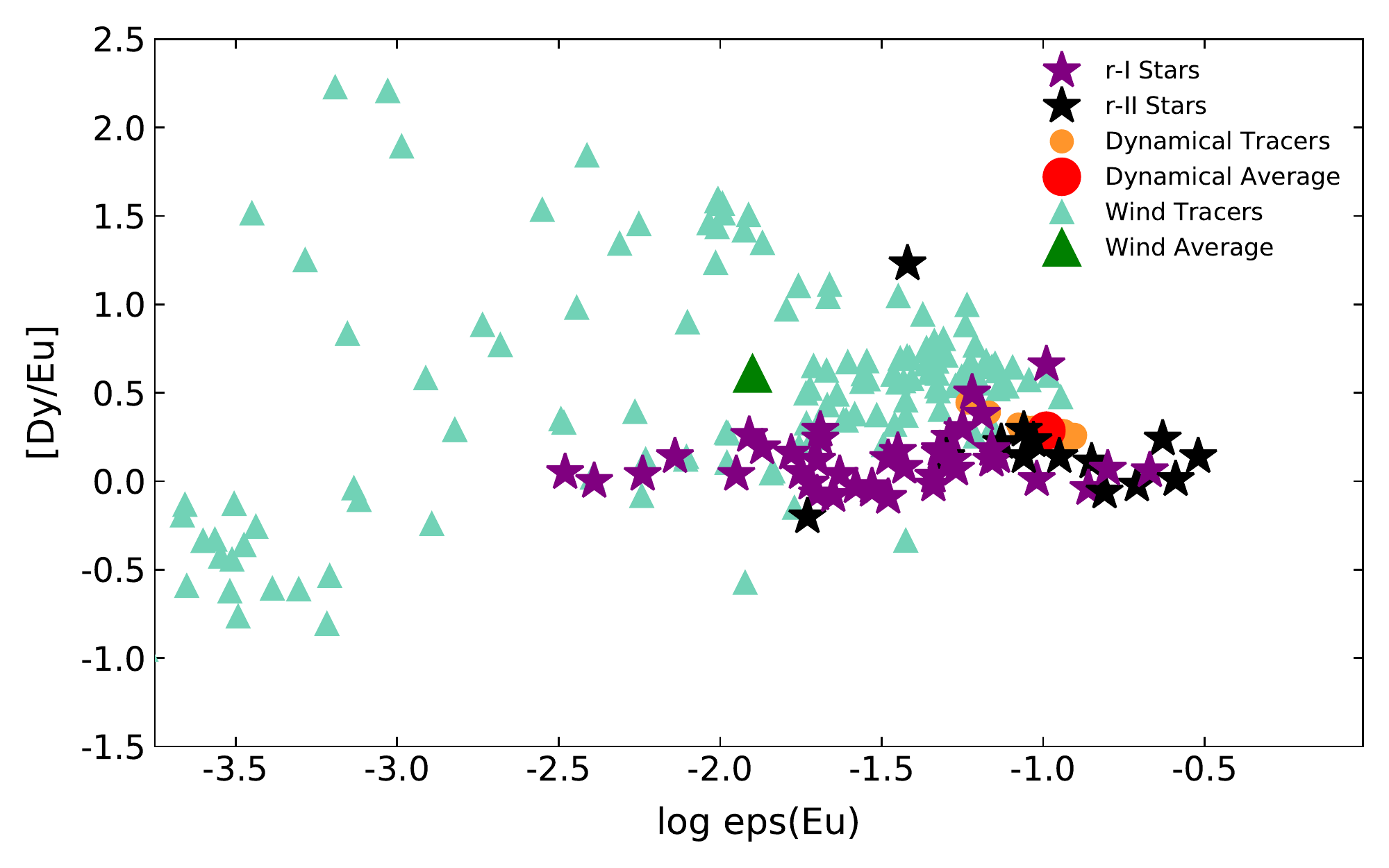} %
     \includegraphics[width=6.1cm]{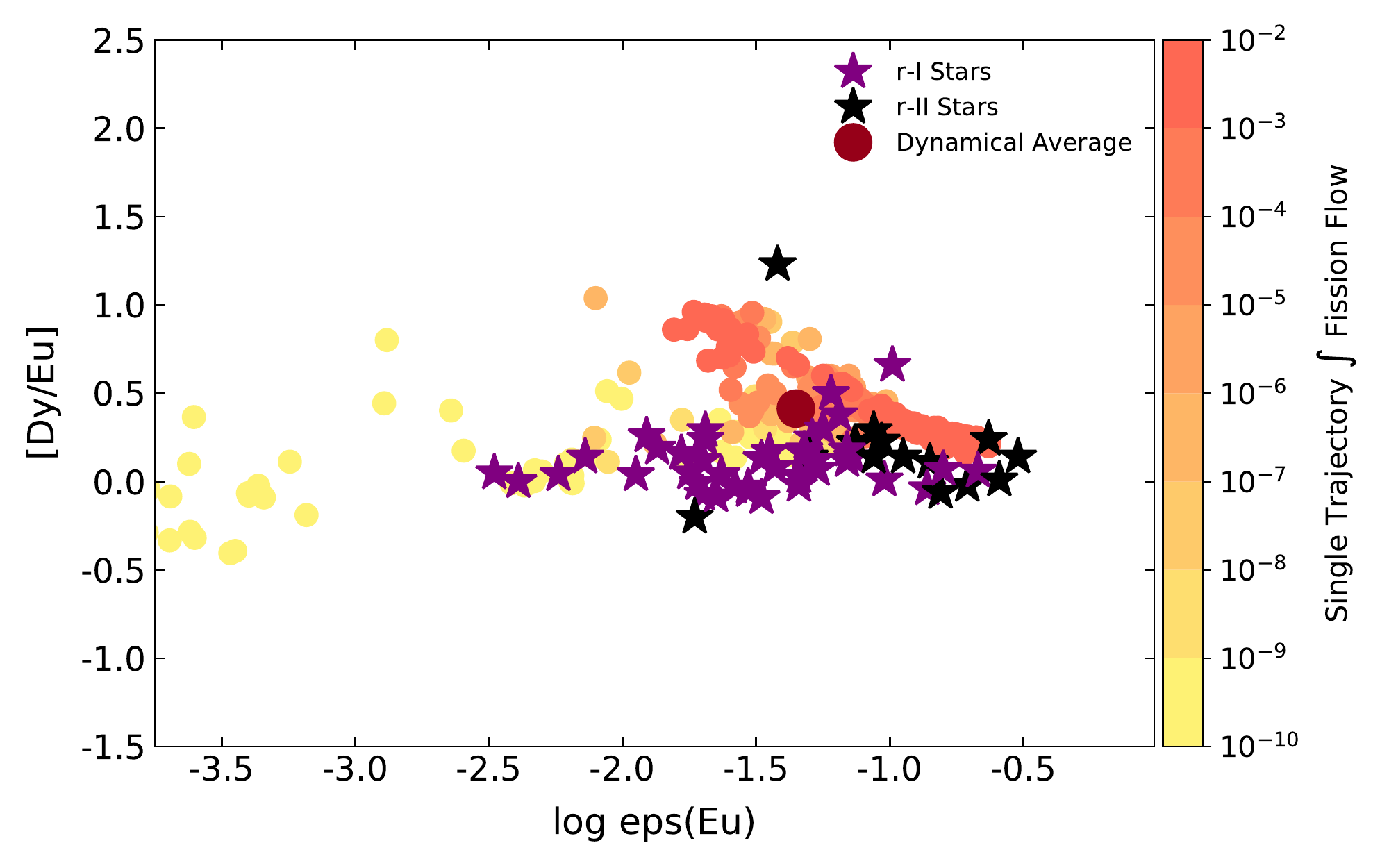}%
    \hspace{0.5cm}
     \includegraphics[width=6.1cm]{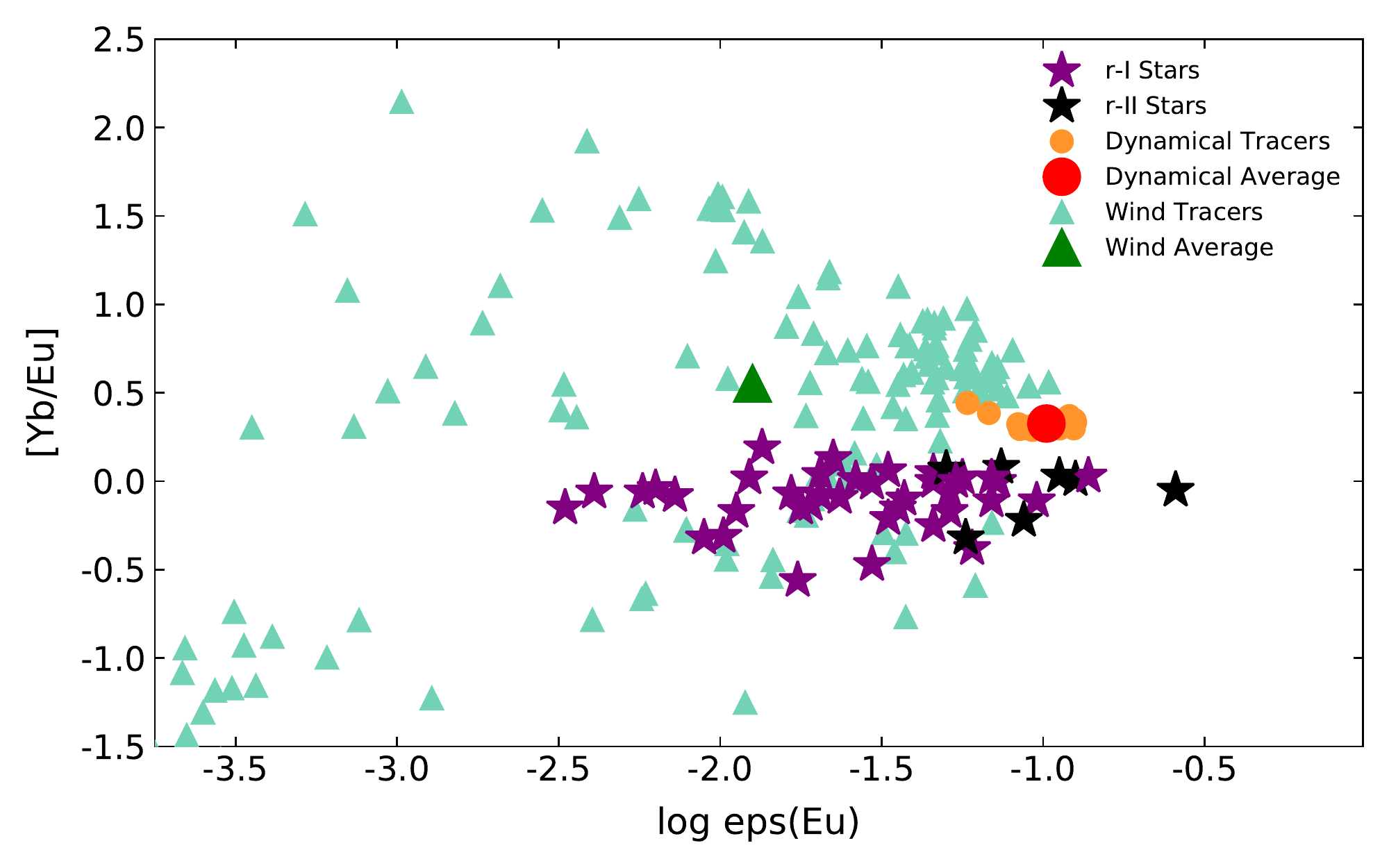} %
     \includegraphics[width=6.1cm]{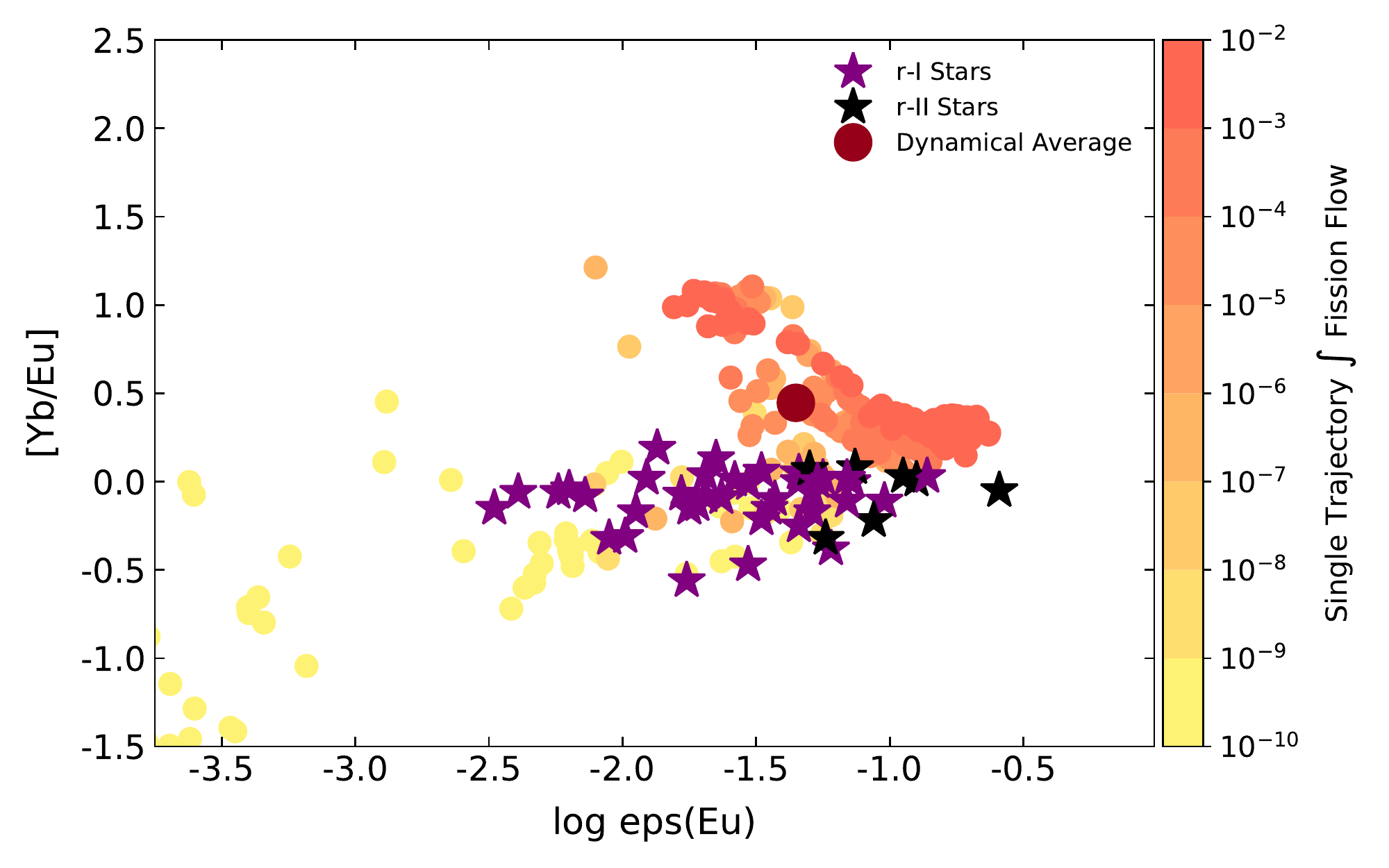}%
  \caption{Same as Fig. 5 but for lanthanide species (La, Pr, Nd, Dy, and Yb) relative to another lanthanide element, Eu (a subset of such plots was shown in \cite{VasshFRLDMrp}).}
\label{fig:secondrIrII}
\end{figure}

In Fig.~\ref{fig:secondrIrII} we examine elemental ratios of some lanthanide elements (La ($Z=57$), Pr ($Z=59$), Nd ($Z=60$), Dy ($Z=66$), and Yb ($Z=70$)) relative to another lanthanide element Eu. The flat trends in the stellar data of r-I and r-II stars show the similarities in the observational data for these ratios and demonstrate the type of trend which is indicative of co-production (as was suggested by the Ag and Pd trends in Fig. \ref{fig:firstrIrII}). Here we see it is not only conditions which reach fissioning nuclei which can accommodate these trends since conditions which instead produce a main $r$ process but do not reach fissioning species also often predict a ratio comparable to the data. Therefore if universality is to be used as a probe of whether fission has played a significant role in setting the final abundances, the elemental ratios of lanthanides relative to each other are not the most relevant feature since fission is not required to accommodate the stellar trends. Rather it is the relative ratios of lanthanides to other regions of the abundance patterns, such as the third peak or light precious metals, such as Ag and Pd, which could reveal the participation of fission in the astrophysical scenario which produced these nuclei.

\section{Conclusions}\label{sec-5}
Fission is a unique probe of the astrophysical origins of the heaviest species observed in nature. Since the participation of this exotic and energetic process implies that elements such as gold have been synthesized prior to the building of the actinides, fission signatures are a means to understand whether an astrophysical event has produced all $r$-process nuclei. Here we discussed the challenges in quantifying the impact of fission in the $r$ process due to the uncertainties in the fission rates and fission fragment distributions of neutron-rich nuclei. We discussed potential signatures of fission which were able to be identified when applying currently available models, such as the heating impact from late-time spontaneous fission on merger light curves. This highlights the need for both experiment and theory to extend their current reach into neutron-rich regions. We showed that recent calculations for fission yields using macroscopic-microscopic theory predict distributions which are distinct from past models via their enhanced width and therefore range of deposition in the $r$ process. We discussed the possible connection between such deposition into the light precious metals and the so-called `robustness' or `universality' of the abundances observed for metal poor stars rich in $r$-process elements. We pointed out that attempts to link fission and universality should be subject to more quantified statements. For instance, the universality observed amongst lanthanide element abundances relative to each other does not require a participation from fission. However universality discussions should be extended to consider the ability of fission to co-produce the light precious metals, such as Ag and Pd, along with the lanthanide elements since fissioning conditions most closely and robustly reproduce [Ag/Eu] and [Pd/Eu] stellar trends. Thus although there remain many mysteries as to the astrophysical origins of heavy elements, including whether or not fission significantly occurs in $r$-process environments, luckily due to the extreme impacts of fission on $r$-process observables, such questions become more approachable in this new era of multi-messenger events and precision observations.

\begin{acknowledgement}
The work of N.V., M.R.M., R.S., and R.V. was partly supported by the Fission In R-process Elements (FIRE) topical collaboration in nuclear theory, funded by the U.S. Department of Energy. Additional support was provided by the U.S. Department of Energy through contract numbers DE-FG02-95-ER40934 (R.S. and T.M.S.), and DE-SC0018232 (SciDAC TEAMS collaboration, R.S. and T.M.S). T.M.S. was supported in part by the Los Alamos National Laboratory Center for Space and Earth Science, which is funded by its Laboratory Directed Research and Development program under project number 20180475DR. R.S. also acknowledges support by the National Science Foundation Hub (N3AS) Grant No. PHY-1630782. The work of R.V. was performed under the auspices of the U.S. Department of Energy by Lawrence Livermore National Laboratory under Contract DE-AC52-07NA27344. M.R.M. was supported by Los Alamos National Laboratory operated by Triad National Security, LLC, for the National Nuclear Security Administration of U.S.\ Department of Energy (Contract No.\ 89233218CNA000001). This work was partially enabled by the National Science Foundation under Grant No. PHY-1430152 (JINA Center for the Evolution of the Elements). This manuscript has been released via Los Alamos National Laboratory report number LA-UR-20-22730.
\end{acknowledgement}


%
%
%

\end{document}